\documentclass[journal]{IEEEtran}

\usepackage{graphicx}
\usepackage{subfigure}
\usepackage{ulem}
\usepackage{cite}
\usepackage{amssymb}
\usepackage{amsmath, amsthm}
\usepackage{xcolor}

\begin{document}

\title{The Role of Criticality of Gene Regulatory Networks in Morphogenesis}

\author{Hyobin~Kim~and~Hiroki~Sayama
\thanks{H. Kim is with the Center for Complexity Sciences (C3), at National Autonomous University of Mexico (UNAM), Coyoac\'an, Mexico City, 04510 Mexico (E-mail: hyobin.kim@c3.unam.mx). H. Sayama is with the Center for Collective Dynamics of Complex Systems, and the Department of Systems Science and Industrial Engineering, at Binghamton University, State University of New York, Binghamton, NY, 13902 USA.}}

\maketitle

\begin{abstract}
Gene regulatory network (GRN)-based morphogenetic models have recently gained an increasing attention. However, the relationship between microscopic properties of intracellular GRNs and macroscopic properties of morphogenetic systems has not been fully understood yet. Here we propose a theoretical morphogenetic model representing an aggregation of cells, and reveal the relationship between criticality of GRNs and morphogenetic pattern formation. In our model, the positions of the cells are determined by spring-mass-damper kinetics. Each cell has an identical Kauffman's $NK$ random Boolean network (RBN) as its GRN. We varied the properties of GRNs from ordered, through critical, to chaotic by adjusting node in-degree $K$. We randomly assigned four cell fates to the attractors of RBNs for cellular behaviors. By comparing diverse morphologies generated in our morphogenetic systems, we investigated what the role of the criticality of GRNs is in forming morphologies. We found that nontrivial spatial patterns were generated most frequently when GRNs were at criticality. Our finding indicates that the criticality of GRNs facilitates the formation of nontrivial morphologies in GRN-based morphogenetic systems. 
\end{abstract}

\begin{IEEEkeywords}
Morphogenetic system, morphogenetic pattern, gene regulaoty network (GRN), random Boolean network (RBN), criticality, cell fate.
\end{IEEEkeywords}

\IEEEpeerreviewmaketitle

\section{Introduction}

\IEEEPARstart{G}{ene} regulatory networks (GRNs) have been an interesting topic from modeling to applications in artificial life and engineering research \cite{jakobi2003harnessing, banzhaf2003artificial, reil1999dynamics, cussat2012controlling, cussat2012using, joachimczak2010evolving, bentley2004adaptive, dellaert1996developmental}. Especially, as a framework to study morphogenesis during developmental processes, many GRN-based morphogenetic systems to form nontrivial morphogenetic patterns or shapes in 2D or 3D space have been actively developed \cite{disset2014toward, schramm2012evolution, knabe2008evolution, schramm2003evolving, joachimczak2008evo, doursat2008programmable, eggenberger1997evolving}. However, the relationship between microscopic properties of intracellular GRNs and macroscopic properties of morphogenetic systems have not been fully explored yet. Thus, we study the relationship between microscopic properties of GRNs and collective properties of morphogenetic systems.

Specifically, we aim to investigate what role the criticality of GRNs plays in morphogenetic pattern formation. The concept of the criticality of GRNs was established by Kauffman \cite{kauffman1996home, kauffman1993origins, kauffman1969metabolic}. He presented a phase transition between ordered, critical, and chaotic regimes in $NK$ random Boolean networks (RBNs) as GRN models. The phase can be varied through parameters such as node in-degree ($K$), internal homogeneity ($p$), and canalizing functions. In an ordered regime, a GRN is highly robust against perturbations. On the contrary, in a chaotic regime, a GRN is too sensitive to perturbations to predict the dynamics. Meanwhile, in a critical regime, a GRN is robust and sensitive at the same time. When perturbations are added to GRNs, critical GRNs conserve existing functions and make new ones simultaneously. That is, the optimal balance between robustness and sensitivity is achieved in a critical regime \cite{aldana2007robustness}. 

By comparing the dynamics of gene expression data of real biological systems with the dynamics of GRN models in ordered, critical, and chaotic regimes, it has been disclosed that the dynamics of biological systems are consistent with the dynamics of GRN models at a critical regime \cite{balleza2008critical, shmulevich2005eukaryotic, shmulevich2004genetic, shmulevich2004activities, aldana2003boolean, kauffman2003random}. Taking a step forward from the relationship between the criticality of GRNs and the dynamics of GRNs in a single cell, we examine the role of the criticality of GRNs in morphogenesis at a collective level.

\begin{figure}[b]
\centering
\includegraphics [width=\columnwidth]{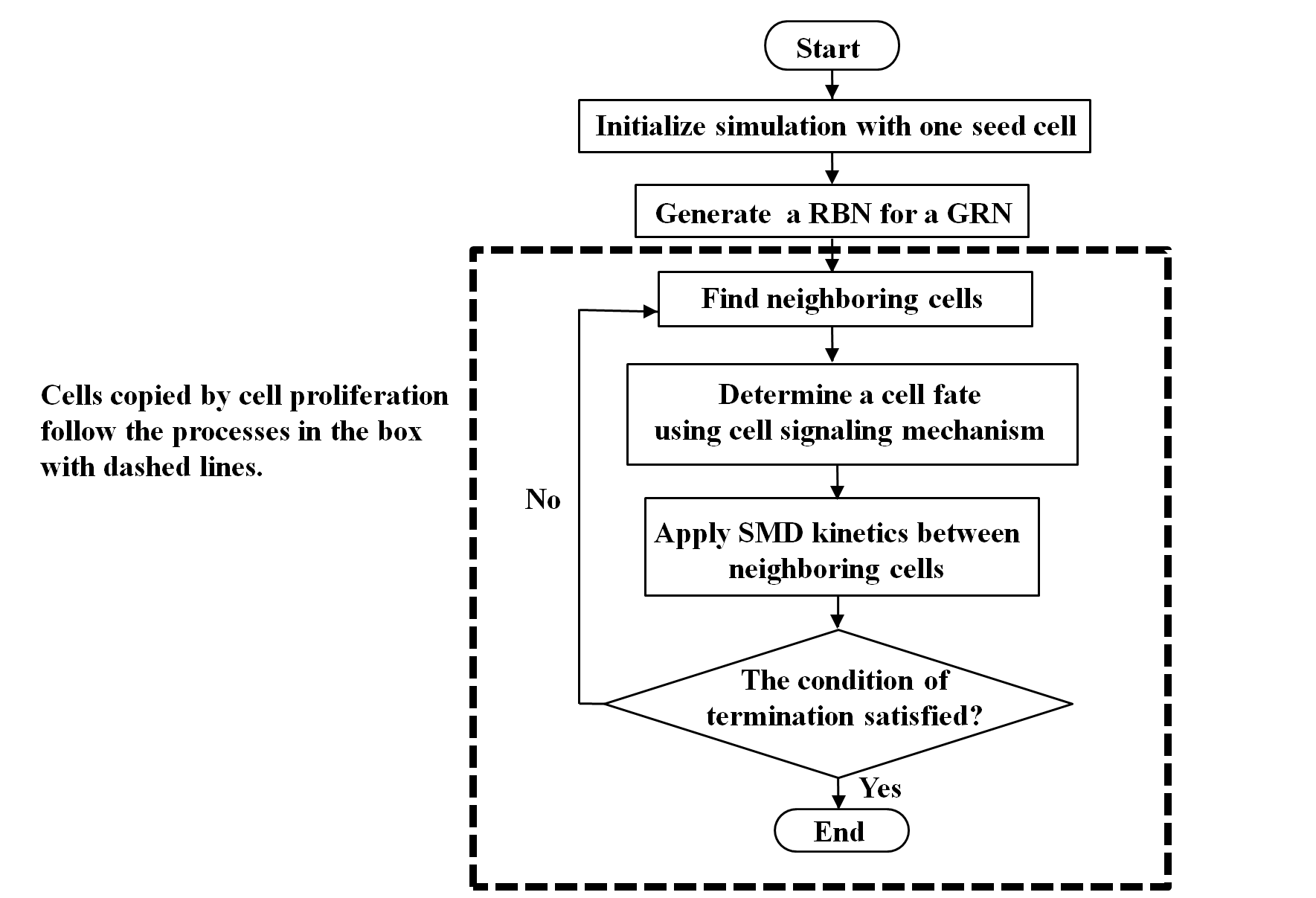}
\caption{Flowchart of simulation in our GRN-based morphogenetic model.}
\label{fig1_flowChart} 
\end{figure}

\section{Model}

Our morphogenetic model starts with one seed cell which has a GRN. The seed cell grows into an aggregation iterating the processes shown in Fig.~\ref{fig1_flowChart} in each time step. In our model, a cell has four fundamental cellular behaviors. If the cell fate is proliferation, the cell divides into two, where the daughter cell is placed within a fixed neighborhood radius ($R$) from the mother cell. In case of apoptosis, the cell dies and disappears. In case of differentiation, the cell is labeled as differentiated. Or in case of quiescence, the cell does not show any cellular behaviors. Cells in a proliferation, differentiation, or quiescence state may switch their fates by cell-cell interactions. The cells' positions in the space are determined by spring-mass-damper (SMD) kinetics. Through these algorithms, diverse morphogenetic patterns are obtained in the model. The simulator was implemented in Java.

\begin{figure*}[t] 
\centering
\subfigure[]{
\includegraphics[width=.8\textwidth]{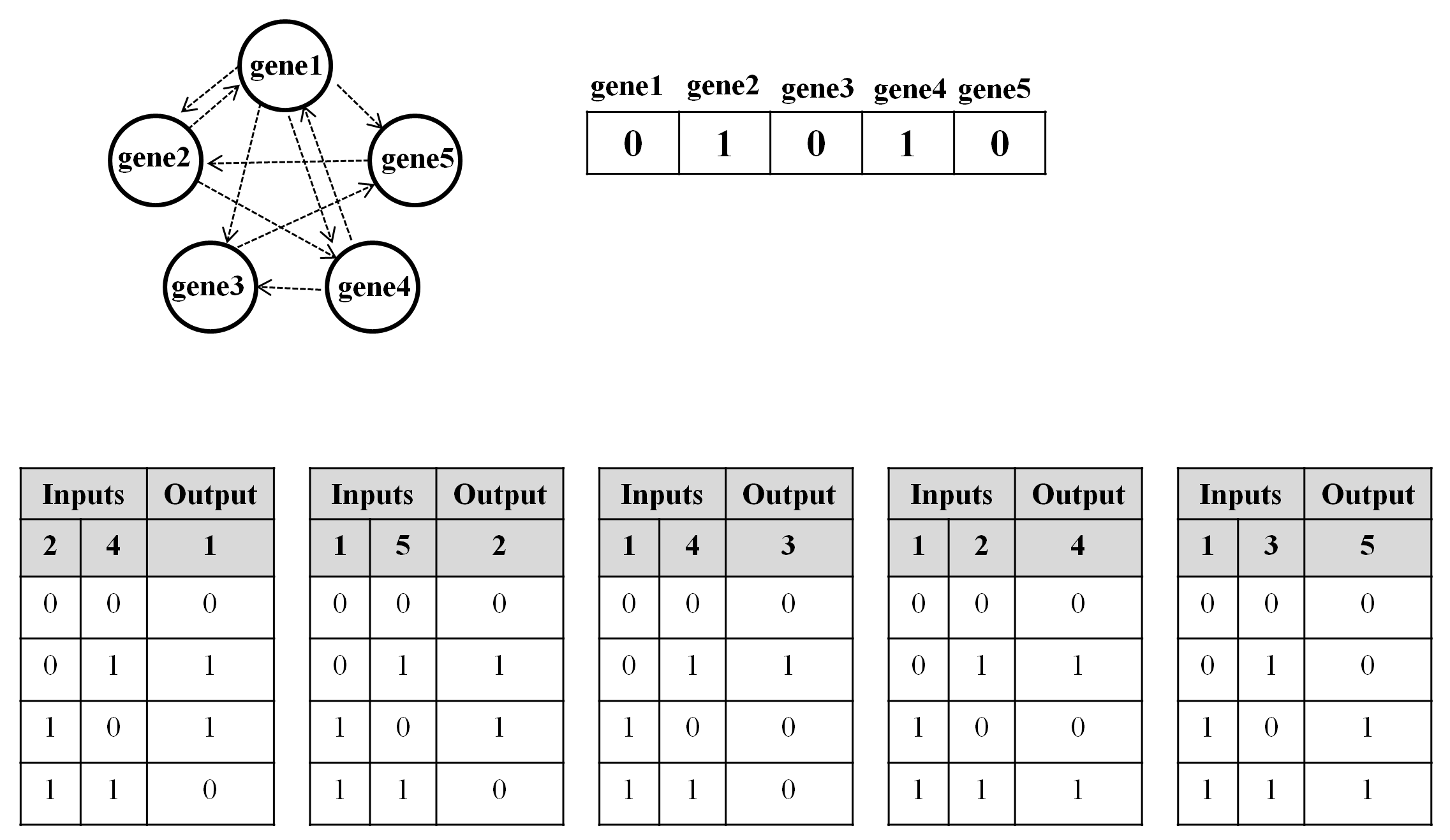}
}
\centering
\subfigure[]{
\includegraphics[width=.45\textwidth,height=6cm]{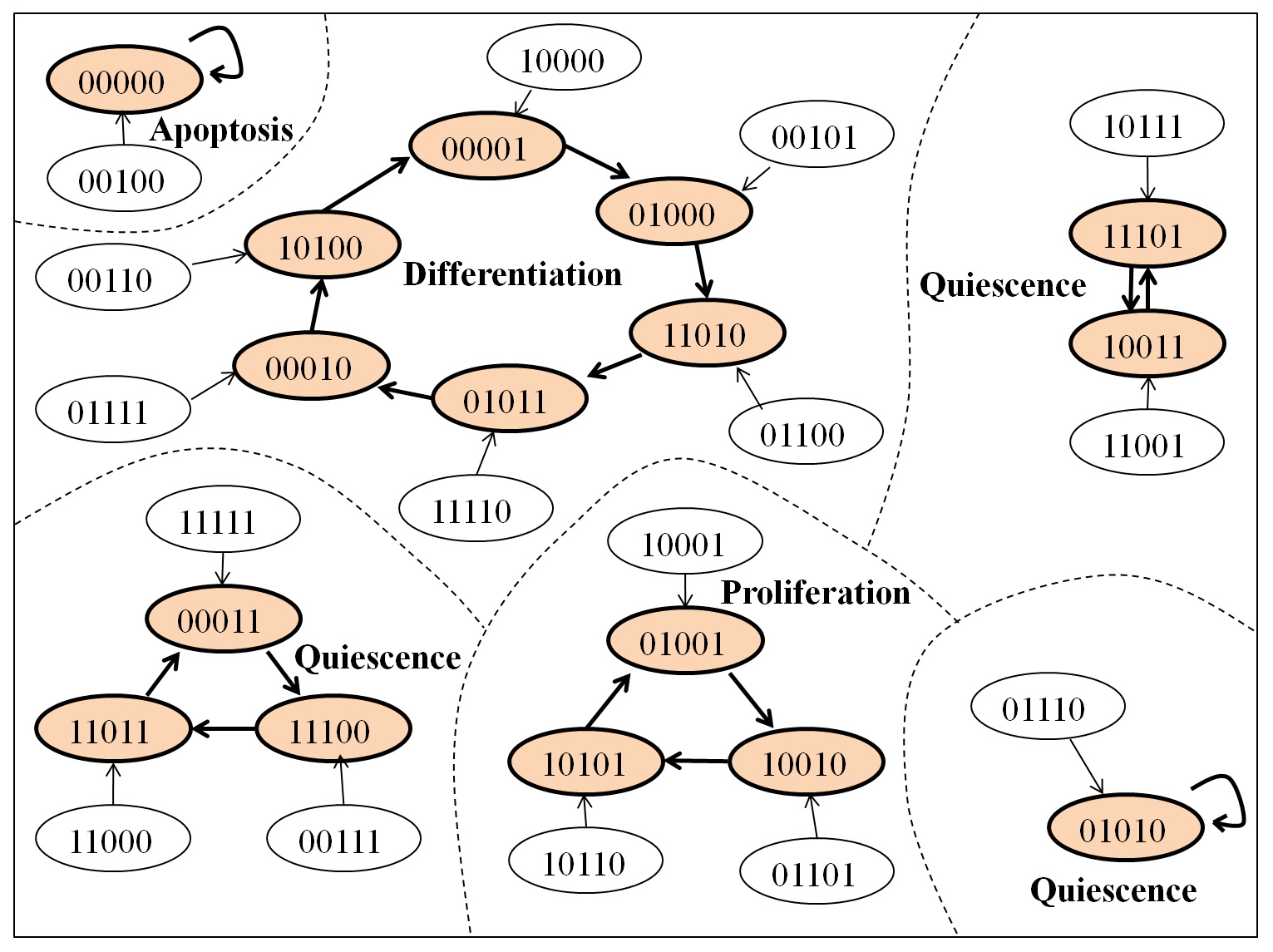}
}
\caption{Schematic diagrams of an example GRN and its state space. (a) A GRN(=RBN) with five nodes (genes) with $K = 2$, and Boolean functions randomly assigned to each node (16 nodes were used in actual simulations). Each node can have either ON (1) or OFF (0). The state of a node is determined by the states of input nodes and assigned Boolean functions. (b) State space of the GRN and randomly assigned four cell fates in it. The state space consists of $2^5 = 32$ configurations and transitions among them. Highlighted are attractors, and the boundaries of their basins of attraction are shown by dashed lines.}
\label{fig2_RBN}
\end{figure*}

\begin{figure*}[t] 
\centering
\subfigure[]{
\includegraphics[width=.7\textwidth,height=4.7cm]{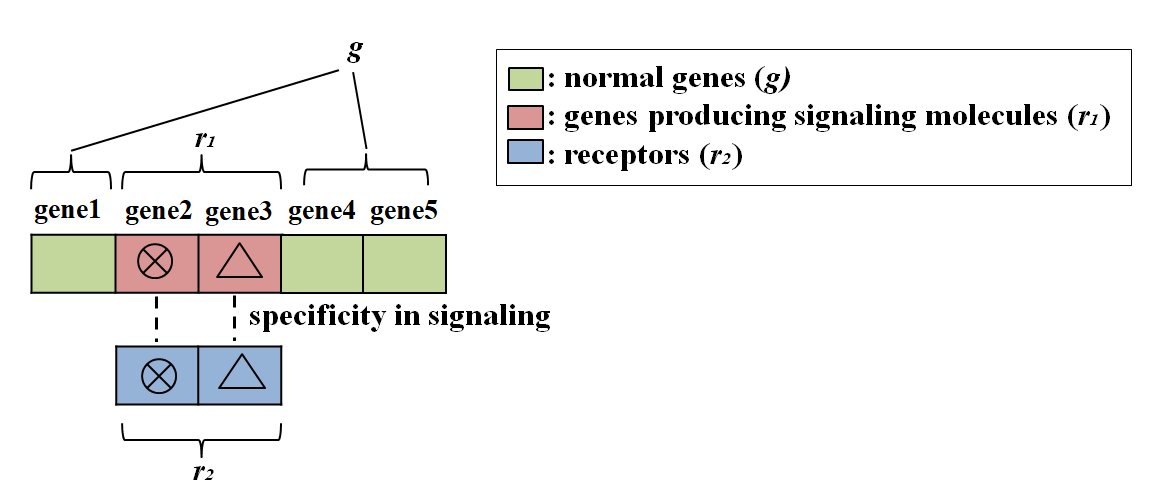}
}
\newline
\subfigure[]{
\includegraphics[width=.9\textwidth,height=7cm]{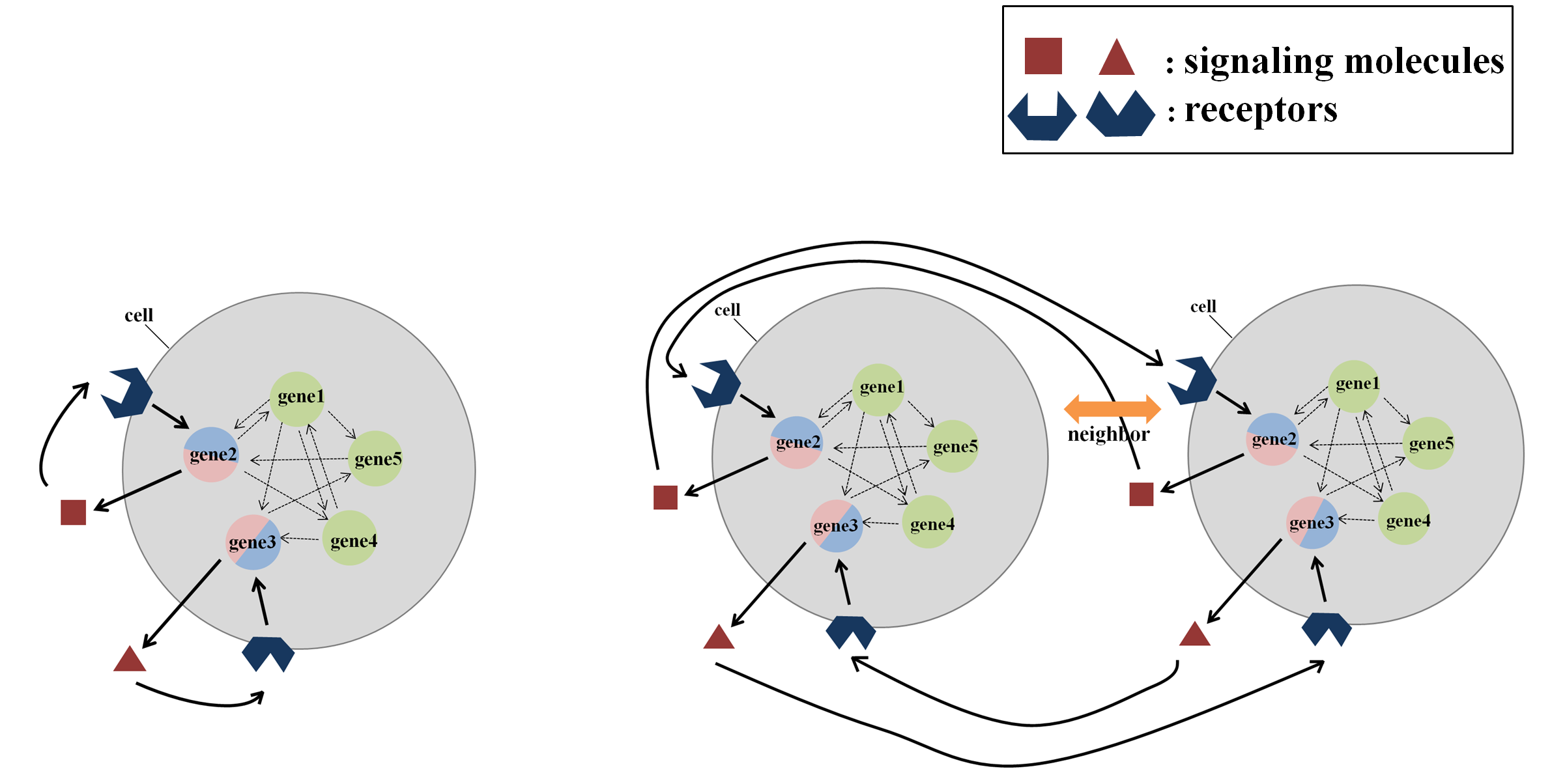}
}
\caption{Cell signaling mechanism for cell-cell interactions. The illustrations explain the concept of cell signaling mechanism. (a) Assignment of genes for cell signaling. g: normal genes. r: special genes for cell signaling. (b) Autocrine (left) and paracrine (right) signaling.}
\label{fig3_cellSignaling}
\end{figure*}

\begin{figure}
\centering
\includegraphics [width=\columnwidth]{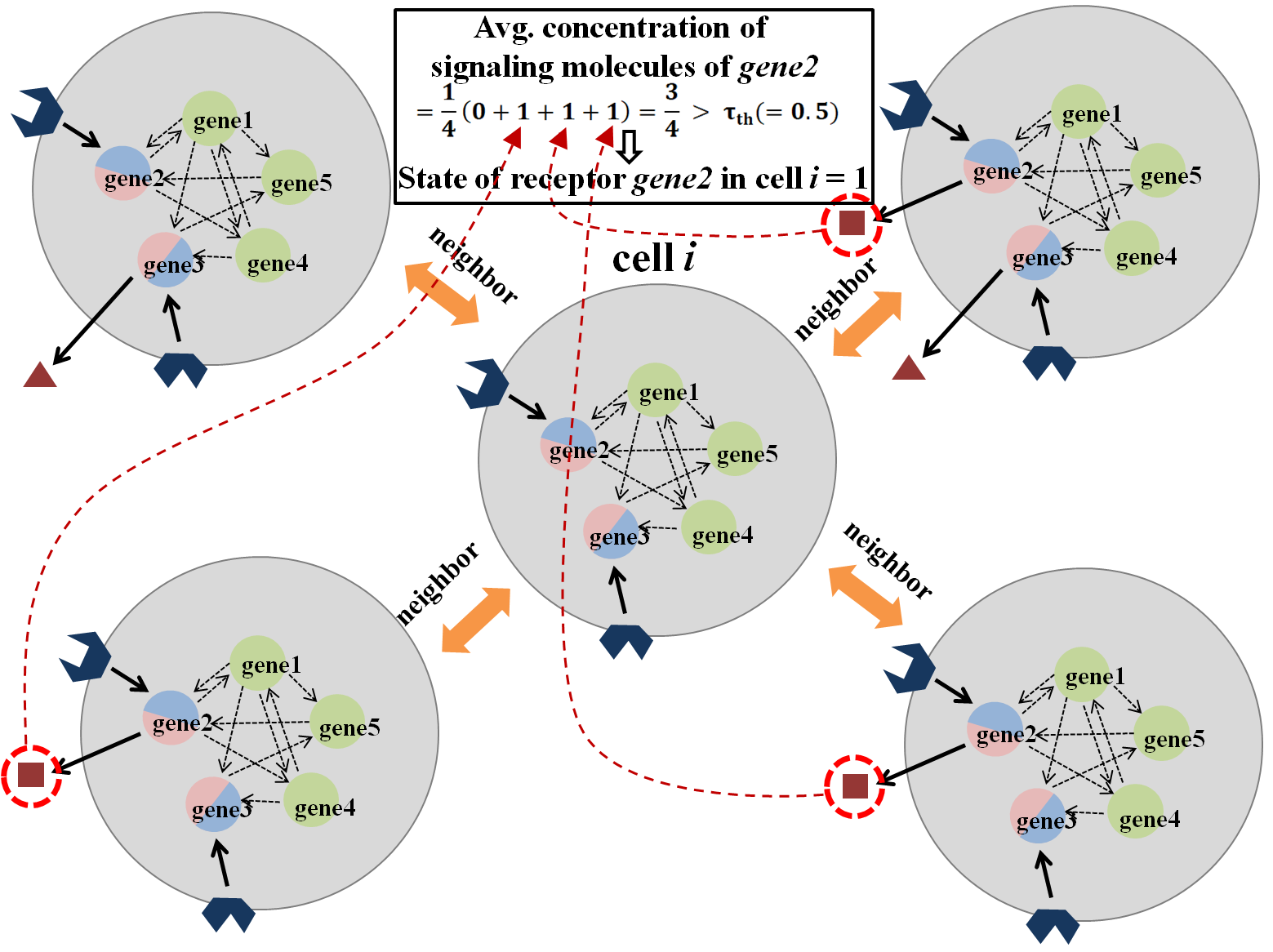}
\caption{A schematic diagram showing how the state of the receptor $gene2$ of cell $i$ is determined by the average concentration of the signaling molecules neighboring cells produce.
\label{fig4_avgConSig}}
\end{figure}

\subsection{Gene Regulatory Network (GRN)}
Our model represents a cell aggregation, where all the cells have an identical $NK$ RBN that consists of 16 nodes ($N$ = 16) as a GRN (Fig.~\ref{fig2_RBN} (a)). As node in-degrees ($K$) of a GRN is varied, the properties of GRNs changes; $K=1$ is ordered, $K=2$ is critical, and $K>2$ is chaotic, on average \cite{kauffman1996home, kauffman1993origins, kauffman1969metabolic}. Based on empirical evidence that attractors of GRNs correspond to cell type/fates, Huang explained stochastic and reversible switching between cell fates using $NK$ Boolean networks \cite{huang2000shape, huang1999gene}. Extending Huang's conceptual framework, we implement $NK$ RBN-based morphogenetic systems. We randomly assign the cell fates to attractors of GRNs in the order of proliferation, differentiation, apoptosis, and quiescence. Quiescence is repeatedly assigned if there are more than four attractors (Fig.~\ref{fig2_RBN} (b)).

\subsection{Switch of Cell Fates by Cell-Cell Interactions}
Transitions between cell fates are caused by perturbations of internal gene expression of a GRN through cell-cell interactions. The cell-cell interactions are based on a cell signaling mechanism of Damiani et al.'s multiple random Boolean networks model on 2D cellular automata \cite{damiani2013modelling, damiani2011cell}. In our model, a GRN of each cell has $n$ genes, which are composed of normal genes ($g$) and special genes ($r$). The special genes are comprised of pairs where genes synthesizing signaling molecules ($r_{1}$) and receptors ($r_{2}$) are matched one to one (Fig.~\ref{fig3_cellSignaling} (a)). This is on the basis of specificity in signaling by which certain signaling molecules can respond to particular receptors. The genes $r_{1}$ synthesize signaling molecules and send messages to other cells within the neighborhood radius ($R$) in the space. Then, the corresponding receptors $r_{2}$ receive the signals by binding to the signaling molecules. 

The cell signaling mechanism is divided into two: autocrine and paracrine (Fig.~\ref{fig3_cellSignaling} (b)). Autocrine is a cell signaling in which receptors are influenced by signaling molecules the cell itself produces when there are no neighboring cells. In contrast, paracrine is a cell signaling where receptors are affected by signaling molecules produced by neighbors. 

The states of the normal genes ($g$) and genes producing signaling molecules ($r_{1}$) are updated by randomly assigned Boolean functions and the states of input nodes. If the states of $r_{1}$ are 1, it means that the genes produce signaling molecules. If the states are 0, signaling molecules are not produced. The states of the receptors $r_{2}$ are updated by the average concentration of the signaling molecules neighboring cells produce. For example, in Fig.~\ref{fig4_avgConSig}, the state of the receptor $gene2$ of cell $i$ is determined by the average concentration of the signaling molecules of the neighboring cells. If the average concentration is larger than a certain threshold ($\tau_{th}$), the state of the receptor $gene2$ becomes activated (1, ON). Otherwise, it becomes inhibited (0, OFF). 

The cell fates are switched through the following steps:
\begin{enumerate}
  \item Check if there are neighboring cells within $R$ or not. If there are neighbors, paracrine signaling is used. Otherwise, autocrine signaling is used.
  \item Determine the states of receptors according to the concentrations of signaling molecules.
  \item Change the states of genes that are connected with the receptors. If the states of the receptors are activated, the states of genes become activated. Otherwise, the states become inhibited.
  \item Determine a cell fate with the attractor the updated gene states finally evolve into over time.
  \item Assign the attractor states as gene states for the next time step.  
\end{enumerate}

\begin{figure}[b]
\centering
\includegraphics [width=0.6\columnwidth]{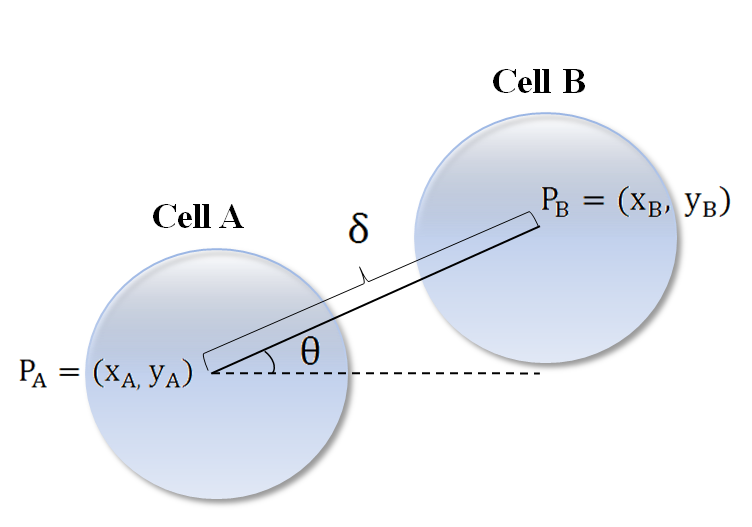}
\caption{Schematic diagram showing the distance and angle between two cells.}
\label{fig5_SMD} 
\end{figure}

\subsection{Spring-Mass-Damper (SMD) Kinetics}
We use spring-mass-damper kinetics for cellular movements following Doursat's approach \cite{doursat2009organically}. Each cell has a position $P = (x,y)$ in a Cartesian coordinate system. Edges connecting cell centers within the neighborhood radius ($R$) are modeled as springs with spring constant $k$ and equilibrium length $l$. For viscous resistance, dampers with damping coefficient $c$ are included. Thus, the equation of movements of a cell is as follows:

$$m\ddot{P}_{AB}=-k(1-\frac{l}{\|P_{AB}\|})P_{AB}-c\dot{P}_{AB}$$
where
$${P_{AB}}={\vec{P}_{B}-\vec{P}_{A}}=(x_{B}-x_{A}, y_{B}-y_{A})=(\delta \cos \theta, \delta \sin \theta),$$
$$\delta={\|P_{AB}\|}, \theta=arctan(\frac{y_{B}-y_{A}}{x_{B}-x_{A}}),$$
$${\|P_{AB}\|}={\|P_{B}-P_{A}\|}=\sqrt{({x_{B}-x_{A}})^2+({y_{B}-y_{A}})^2}$$

Fig.~\ref{fig5_SMD} visually shows the mathematical quantities of $\delta$ (the Euclidean distance between two cells) and $\theta$ (the angle between two cells). Here, we neglect the effect of inertia. That is, we replace $m\ddot{P}_{AB}$ with zero. Then, the equation for a position update is the following at each time step $\Delta$t = 1:\\
$$\Delta{P_{B}}=-\Delta{P_{A}}=\frac{\Delta{P_{AB}}}{2}=\frac{-k}{2c}(1-\frac{l}{\|P_{AB}\|}){P_{AB}}$$

The position updating rule allows physical interactions such as pushing, adhesion, and movements among neighboring cells within $R$. 

To obtain diverse shapes of spatial patterns, we determine the values of parameters $k, l$, and $c$ depending on the cell fates and add perturbations to the position $(x, y)$ values. For the dependence of parameters $k, l$ and $c$ on cell fates, all the possible cell fates ([$\alpha$-$\beta$]) between two cells are categorized into six types: [\textit{proli-proli}], [\textit{proli-diff}], [\textit{proli-qui}], [\textit{diff-qui}], [\textit{diff-diff}], and [\textit{qui-qui}], where \textit{proli} is proliferation, \textit{diff} is differentiation, and \textit{qui} is quiescence. Cells disappearing in the space due to apoptosis are not included. Thus, $k, l$, and $c$ can take six different sets of values according to the cell fates. All the eighteen values (six $k_{[\alpha-\beta]}$ values, six $l_{[\alpha-\beta]}$ values, and six $c_{[\alpha-\beta]}$ values) are randomly chosen in certain ranges in each simulation run ($k, l$ and $c$ in TABLE ~\ref{table:parameters}). In the case of the perturbations, we add small perturbation values to the updated coordinates.
  
By introducing the dependence of $k, l$ and $c$ on cell fates and perturbations to the position $(x, y)$ values, the final position of cell $A$ having cell $B$ as its neighboring cell is the following:
 
$${P}_{A}(t+1)={P}_{A}(t)+(\Delta{P}_{A})_{[\alpha-\beta]}+\omega_{[\alpha-\beta]}$$
where $\alpha$ is cell $A$'s cell fate and $\beta$ is cell $B$'s cell fate. $\omega$ is the perturbation to the updated coordinate of cell $A$.

\section{Experiments}
We conducted 10,000 independent computational simulations of morphogenetic cell growth processes to see if there were any significant differences among the four groups ($K = 1, 2, 3, 4$). Specifications of parameters for the simulations were as follows:\\
\begin{itemize}
  \item[$-$] Cells were placed in a two dimensional 700 $\times$ 700 (in arbitrary unit) square area.
  \item[$-$] In each run, the cell population growth was limited up to 200 to keep computational loads reasonable.
  \item[$-$] The simulations were terminated when the time step ($t$) was 1,000 or there existed no cell in the space because of apoptosis. 
  \item[$-$] The values of parameters regarding GRNs, cell-cell interactions, and SMD kinetics are shown in TABLE ~\ref{table:parameters}.
\end{itemize}

\begin{table}[htbp]
\caption{Parameters and their values for simulations}
\begin{center}
\tabcolsep=0.2cm
\begin{tabular}{ll} 
 \hline
 \bf Parameter & \bf Value \\ 
 \hline
 Number of nodes ($N$) & 16 \\ 
 Number of in-degrees per node ($K$)  & 1, 2, 3, 4 \\ 
 Neighborhood radius ($R$) & 30 \\
 Number of special genes ($r$) & 2 \\
 Threshold of signaling molecules ($\tau _{th}$) & 0.5 \\
 Spring constant ($k$) & $k$ $\in$ unif(0, 1) $\subset$ $\mathbb{R}$\\
 Spring equilibrium length ($l$) & $l$ $\in$ unif(0, 100) $\subset$ $\mathbb{R}$\\
 Damper coefficient ($c$) & $c$ $\in$ unif(0, 200) $\subset$ $\mathbb{R}$\\ 
 \hline
\end{tabular}
\end{center}
\label{table:parameters}
\end{table}

\subsection{Measures for Morphogenetic Patterns}
\begin{figure} [b]
\centering
\subfigure[]{
\includegraphics[width=.3\columnwidth,height=2.2cm]{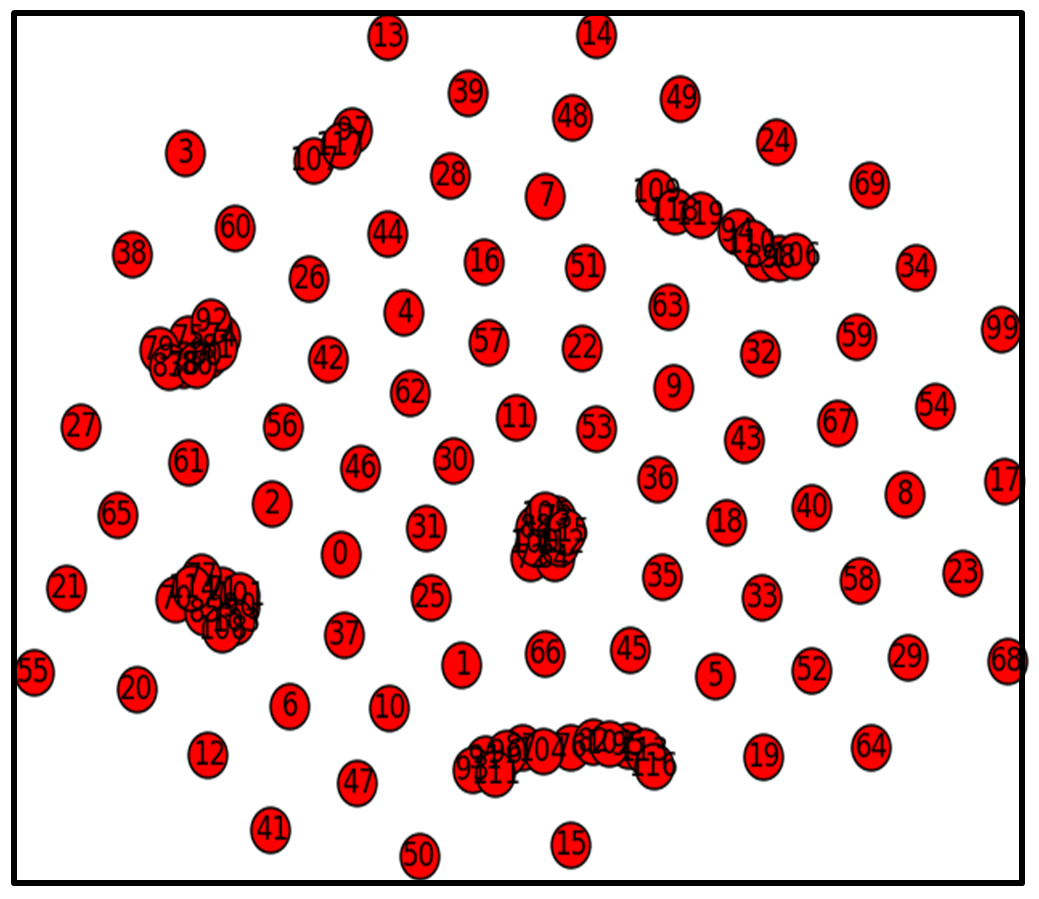}
}
\subfigure[]{
\includegraphics[width=.3\columnwidth,height=2.2cm]{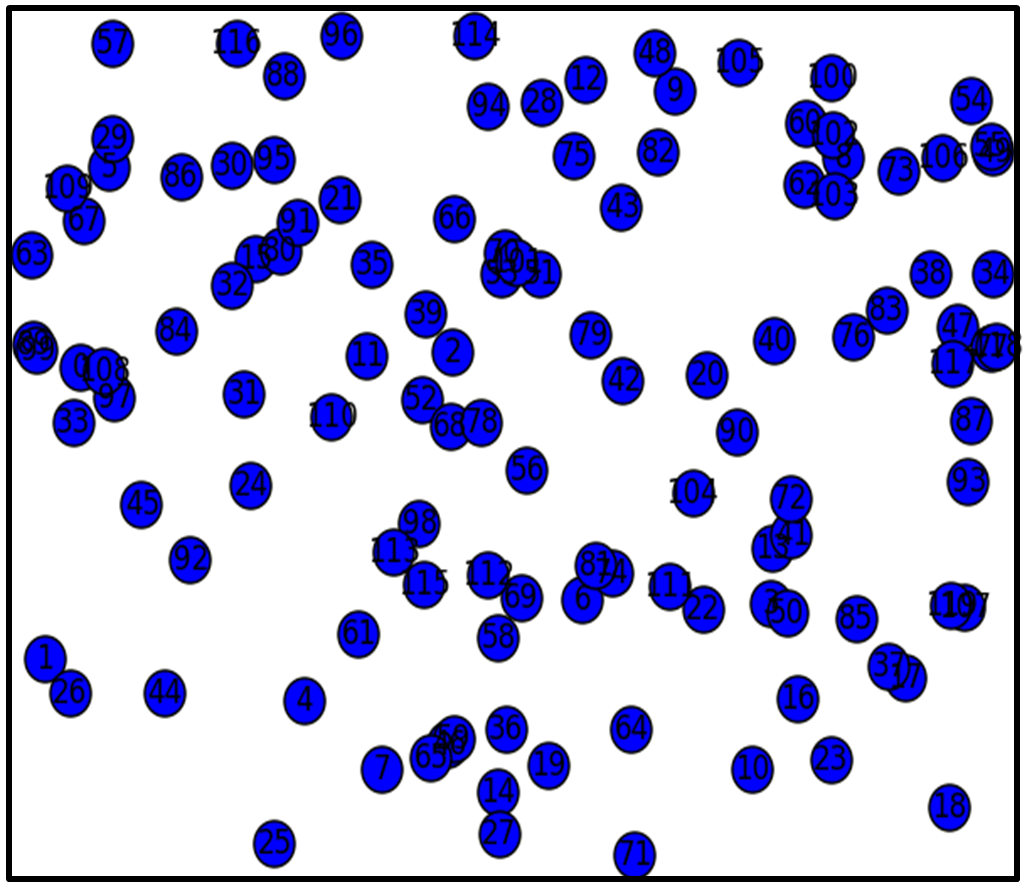}
}\\
\centering
\subfigure[]{
\includegraphics[width=.6\columnwidth,height=2.5cm]{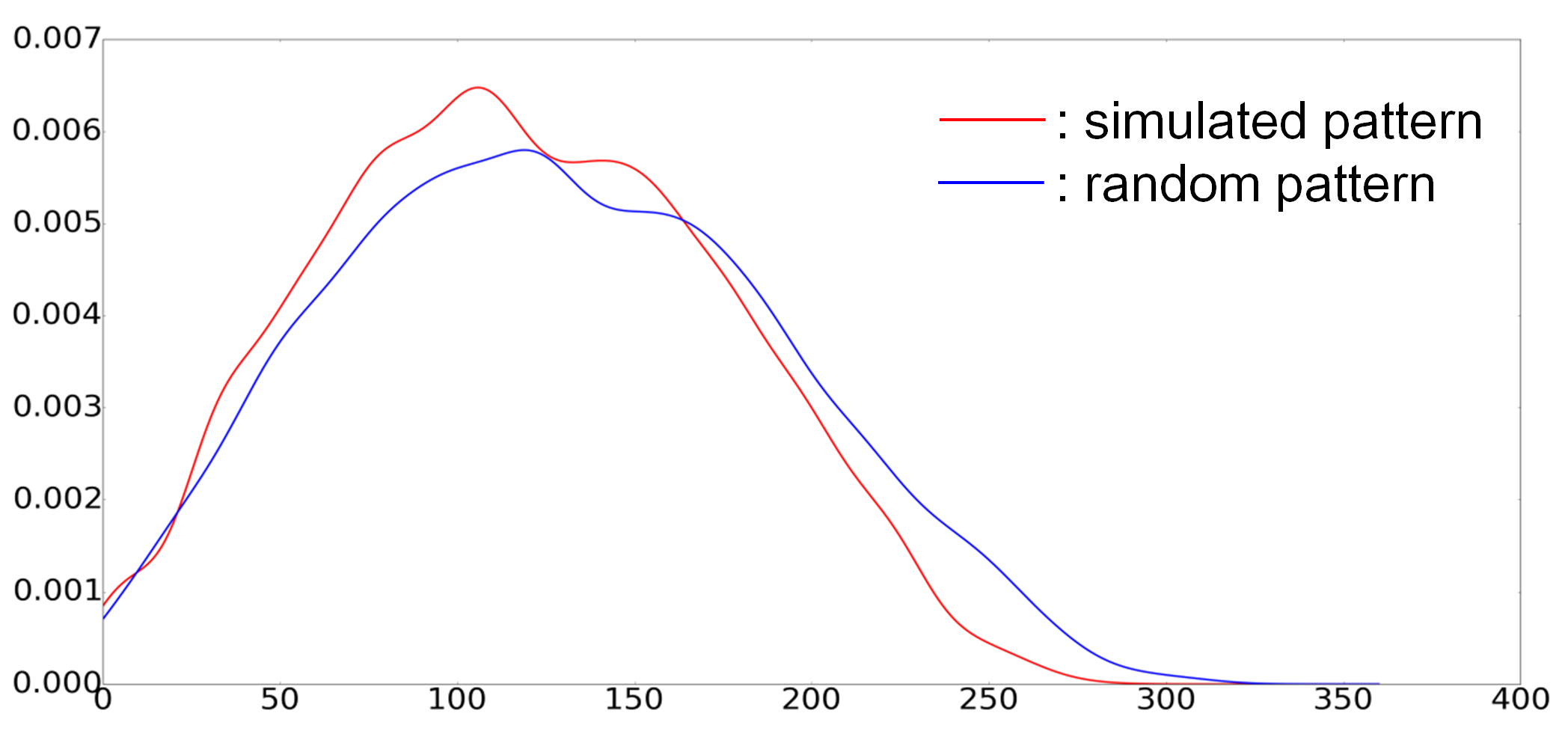}
}
\caption{Comparison of patterns using KL divergence. (a) Morphogenetic pattern obtained from a simulation. (b) Random pattern from a uniform distribution. (c) Distribution curves of pairwise particle distance measurements for simulated and random patterns. Probability density functions of each curve are estimated by Gaussian kernel density estimation.}
\label{fig6_KLD} 
\end{figure}

\begin{figure} [t]
\centering
\includegraphics [width=0.8\columnwidth]{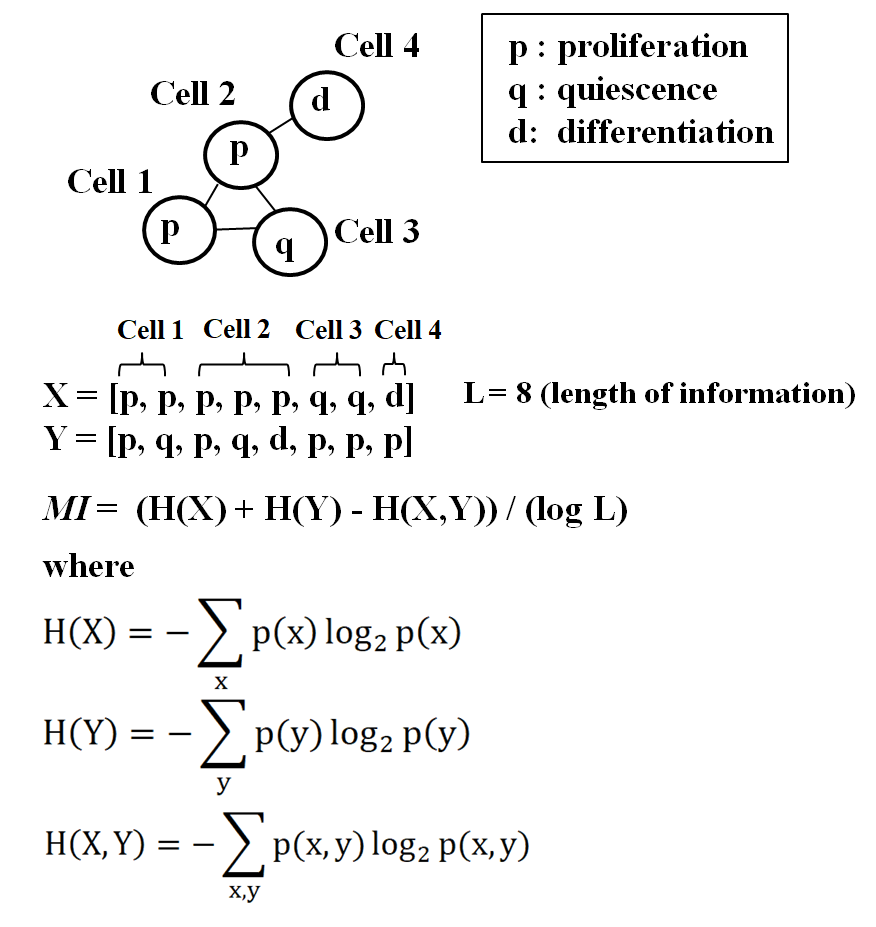}
\caption{An example for the calculation of mutual information between cell fates of cells and their neighboring cells. The calculated mutual information was divided by log L for normalization purposes.}
\label{fig7_MI} 
\end{figure}

We obtained a spatial pattern for each independent simulation run. The following properties were measured from the cells' positions and states based on our previous approaches \cite{sayama2014four, sayama2011quantifying}. Here all the measures were acquired from the final configuration of each simulation.
\begin{itemize}
  \item \textbf{Number of cells} (\textit{numOfCells}). The total number of cells was counted in a morphogenetic pattern.
  \item \textbf{Average distance of cells from center of mass} (\textit{massDistance}). Euclidean distances were calculated from each cell position to the center of mass (i.e., the point with the average coordinates of all the cells.
  \item \textbf{Average pairwise distance} (\textit{pairDistance}). Euclidean distances were measured from two randomly sampled cells' positions. For the average, 10,000 pairs were sampled with replacement.
  \item \textbf{Kullback-Leibler divergence between pairwise particle distance distributions of morphologies} (\textit{kld}). To detect nontrivial patterns, the Kullback-Leibler (KL) divergence between pairwise particle distance distributions of a morphogenetic pattern and a random pattern were measured. Specifically, a pair of coordinates of cells were randomly sampled 10,000 times to generate an approximate pairwise particle distance distribution (Fig.~\ref{fig6_KLD} (c)), first from the morphogenetic pattern (Fig.~\ref{fig6_KLD} (a)), and then from a randomly distributed pattern made of the same number of cells within the same spatial dimensions (Fig.~\ref{fig6_KLD} (b)). 
  \item \textbf{Mutual information between cell fates of cells and their neighboring cells} (\textit{MI}). To examine how much informational correlation exists between the fate of a cell and that of its neighbors, mutual information (\textit{MI}) was calculated. Fig.~\ref{fig7_MI} is an example showing how to calculate \textit{MI} in a morphogenetic pattern. X is a set of cell fates from cell 1 to cell 4 (The cell fate was repeatedly written depending on the number of its neighbors.), and Y is a set of those of their neighboring cells. \textit{MI} is calculated from the marginal entropies of X and Y (H(X), H(Y)), and the joint entropy of X and Y (H(X,Y)). The larger \textit{MI} is, the more strongly correlated with the fate of its neighboring cells the fate of a cell is. When there was only one cell, \textit{MI} was set to 0.\\
\end{itemize}

\begin{figure} 
\centering
\subfigure[]{
\includegraphics[width=.4\columnwidth,height=2.5cm]{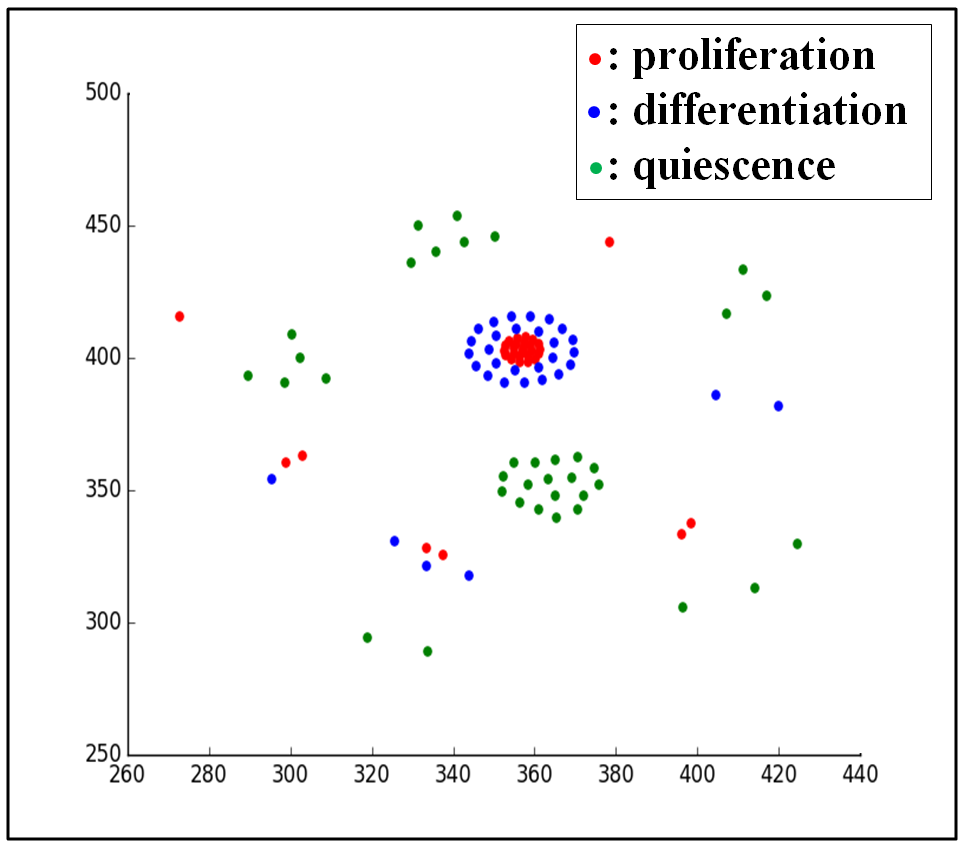}
}
\subfigure[]{
\includegraphics[width=.4\columnwidth,height=2.5cm]{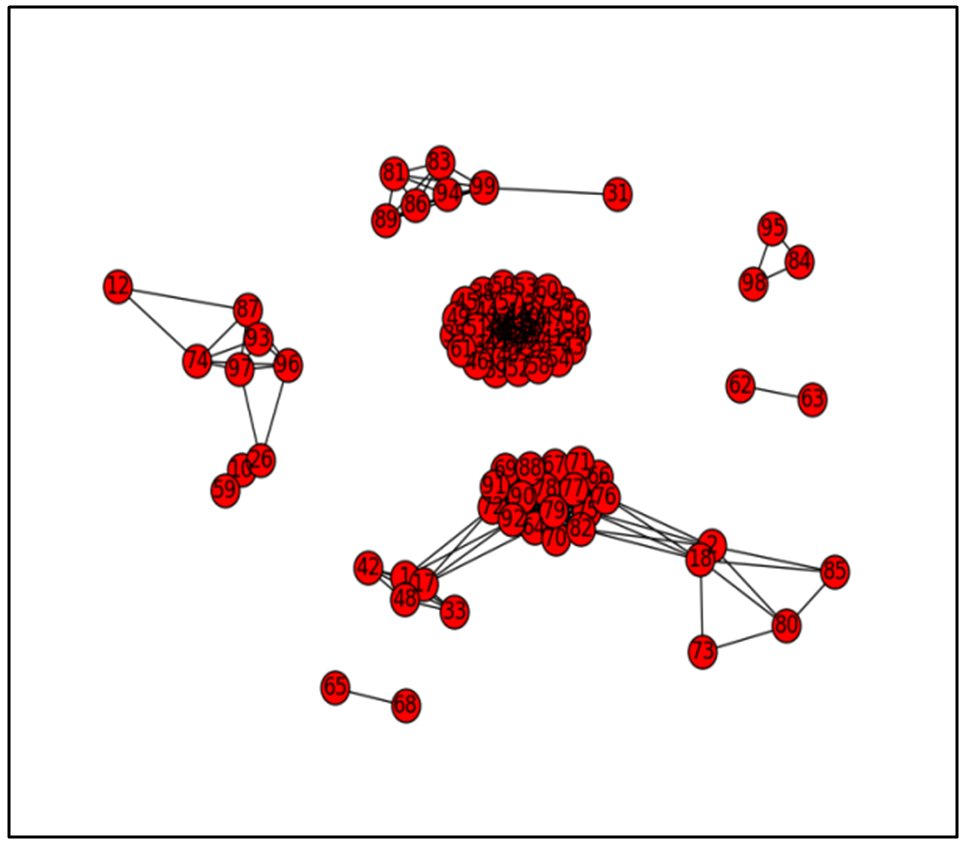}
}\\
\caption{Network construction for morphogenetic pattern analysis. (a) Original morphogenetic pattern snapshot. (b) Network constructed from cells’ positions in (a).}
\label{fig8_netConstruction} 
\end{figure}

In addition, topological properties of the morphogenetic patterns were measured by constructing a network from each morphology. Specifically, each cell was connected to other cells within the neighborhood radius ($R$) in the space. This method is a simpler network construction process than our previous approach \cite{sayama2014four}. Fig.~\ref{fig8_netConstruction} is an example showing an original morphogenetic pattern and a network constructed using the network construction process from it.\\
\begin{itemize}
  \item \textbf{Number of connected components} (\textit{numConnComp}). In a constructed network, a connected component refers to a subgraph where there exists a path between every pair of nodes. A single isolated cell was considered one connected component by itself.
  \item \textbf{Average size of connected components} (\textit{meanSizeConnComp}). The size of a connected component is the number of nodes in it. In a network, the mean of sizes of connected components was measured. In the case that there was no connected component, the value was set to 0.
  \item \textbf{Homogeneity of sizes of connected components} (\textit{homoSizeConnComp}). This examines how similar the sizes of connected components are in a network. It was measured as one minus the normalized entropy in the distribution of sizes of connected components. In the case that there was only one connected component, the value was set to 1.
  \item \textbf{Size of the largest connected component} (\textit{sizeLarConnComp}). This refers to the maximum size of the connected components in a network.
  \item \textbf{Average size of connected components smaller than the largest one} (\textit{meanSizeSmaller}). Except for the largest connected components, the mean of sizes of connected components was calculated in a network. In the case that there was only one connected component, the value was set to 0.
  \item \textbf{Average clustering coefficient} (\textit{avgCluster}). This describes how densely connected the nodes are to each other in a network.
  \item \textbf{Link density} (\textit{linkDensity}). This  quantifies the density of connections in a network.\\
\end{itemize}

From each simulation run, we obtained the values of 12 measures above. In the case that there was no cell, all the values of the measures for morphogenetic patterns were set to 0.

\subsection{Measures to investigate the relationship between GRNs and Expressed Cell Fates}
To investigate the relationship between the criticality of GRNs and the cell states, the following properties were measured from the sizes of basins of a GRN and cells' fates of a morphogenetic pattern:\\
\begin{itemize}
  \item \textbf{Basin entropy.} Basin entropy which was suggested by Krawitz measures the complexity of information that a system is capable of storing as follows \cite{krawitz2007basin}:
$$H_{basin} = -\sum_{\rho} P_{\rho} \cdot \log_2 P_{\rho}$$
where the weight $P_{\rho}$ of  an attractor is the size of the basin of the attractor $\rho$, divided by the size of the state space ($2^N$) of a GRN. Hence, 
$$\sum_{\rho} P_{\rho} = 1$$
In the context of GRNs, the basin entropy represents the effective functional versatility of the cell. In our model, we measured basin entropy based on the basin sizes of attractors to which each cell fate was assigned. We observed the versatility of four cellular functions (proliferation, differentiation, apoptosis, quiescence). 
For example, the basin entropy value in Fig.~\ref{fig2_RBN} (b) is as follows:
$$H_{basin} = - P_{\textit{proli}} \cdot \log_2 P_{\textit{proli}} - P_{\textit{diff}} \cdot \log_2 P_{\textit{diff}}$$
$$ - P_{\textit{apop}} \cdot \log_2 P_{\textit{apop}} - P_{\textit{qui}} \cdot \log_2 P_{\textit{qui}}$$
$$= -\frac{6}{2^5} \cdot \log_2 \frac{6}{2^5} - \frac{12}{2^5} \cdot \log_2 \frac{12}{2^5} $$
$$ - \frac{2}{2^5} \cdot \log_2 \frac{2}{2^5} - \frac{12}{2^5} \cdot \log_2 \frac{12}{2^5} = 1.764 $$
where \textit{proli} is proliferation, \textit{diff} is differentiation, \textit{apop} is apoptosis, and \textit{qui} is quiescence.

\item \textbf{Cell fates entropy.} Similarly, cell fates entropy was measured as follows: 
$$H_{cell fates} = -\sum_{f} P_{f} \cdot \log_2 P_{f}$$
where $P_{f}$ is the number of cells expressing a cellular function $f$ (proliferation, differentiation, apoptosis, quiescence), divided by the total number of all the cells at the end of each simulation. Thus, 
$$\sum_{f} P_{f} = 1$$
In the case that there were no cells expressing proliferation (differentiation/ apoptosis/ quiescence), its log value was set to 0.
\end{itemize}

\begin{figure}[b]
\centering
\includegraphics [width=0.8\columnwidth]{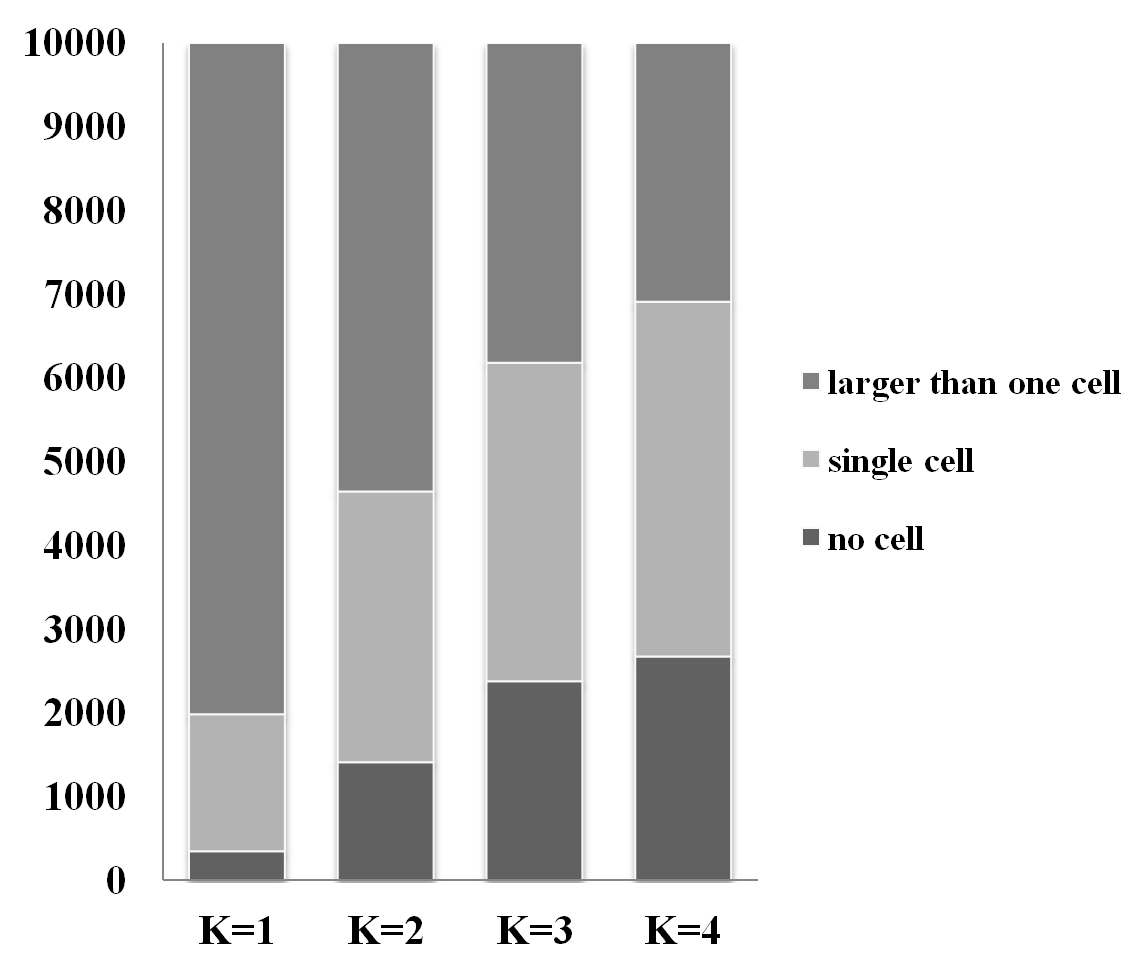}
\caption{Distributions of morphogenetic patterns according to the number of cells for $K = 1, 2, 3, 4$.}
\label{fig9_patternDist} 
\end{figure}

\begin{figure*}[htbp] 
\centering
\subfigure[$K = 1$.]{
\includegraphics[width=\textwidth,height=3.8cm]{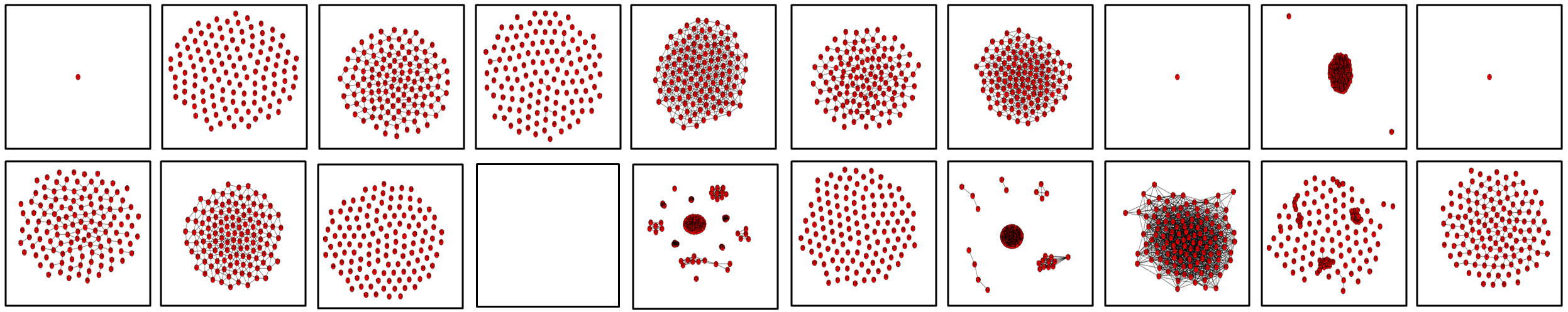}
}
\centering
\subfigure[$K = 2$.]{
\includegraphics[width=\textwidth,height=3.8cm]{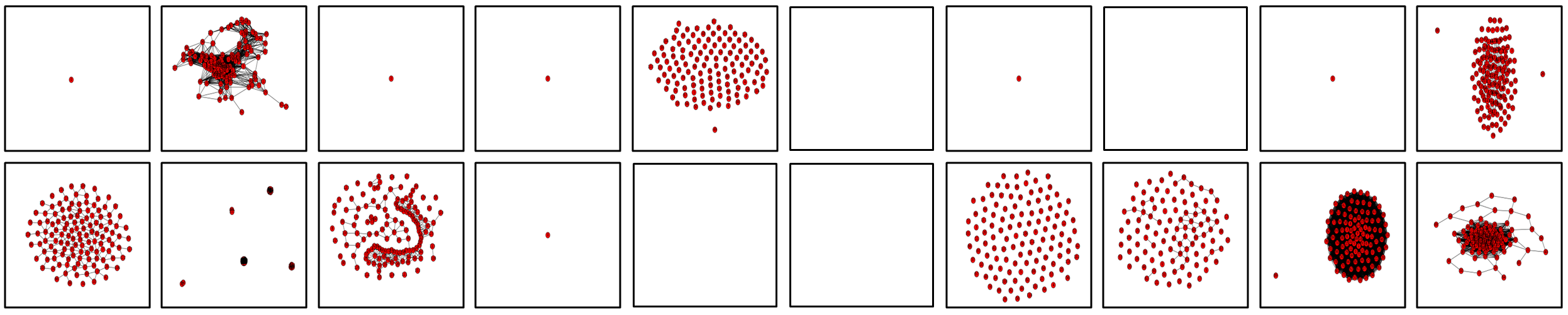}
}
\centering
\subfigure[$K = 3$.]{
\includegraphics[width=\textwidth,height=3.8cm]{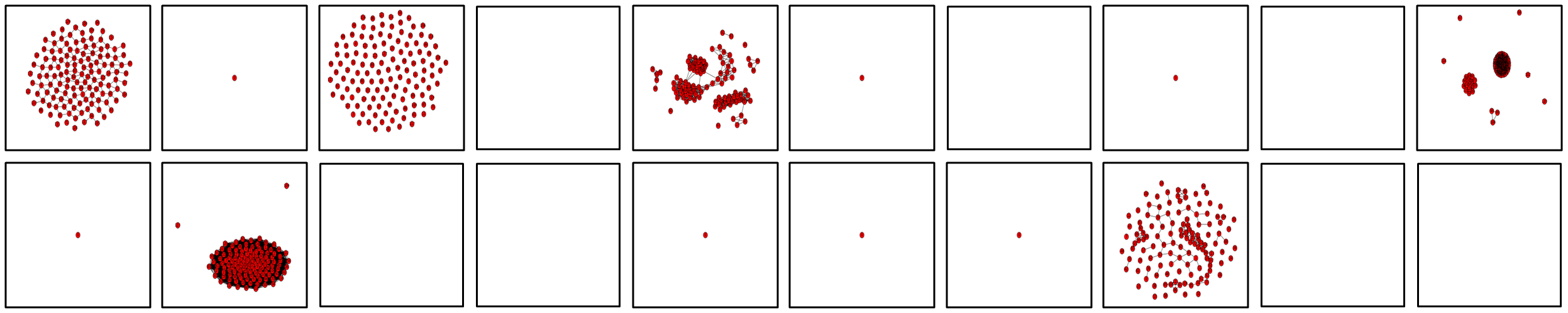}
}
\centering
\subfigure[$K = 4$.]{
\includegraphics[width=\textwidth,height=3.8cm]{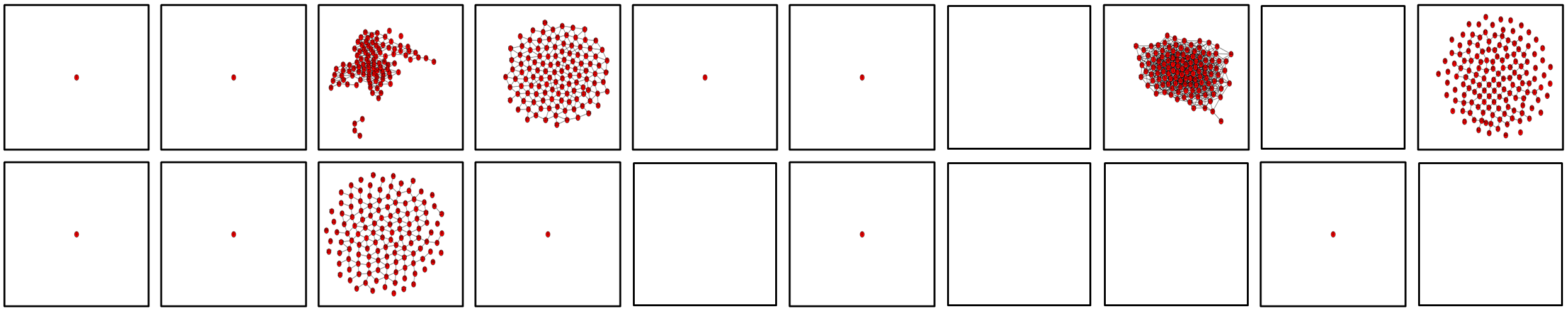}
}
\caption{Different morphogenetic patterns represented with networks for $K = 1, 2, 3, 4$. The patterns are acquired from 20 randomly sampled simulations. (a) $K = 1$. (b) $K = 2$. (c) $K = 3$. (d) $K = 4$.}
\label{fig10_patterns}
\end{figure*}

\section{Results \& Discussion}
Fig.~\ref{fig9_patternDist} shows distributions of the morphogenetic patterns based on the number of cells at the end of each simulation: \textit{larger than one cell, single cell}, and \textit{no cell}. We found that the larger $K$ is, the more frequent the cases of \textit{no cell} and \textit{single cell} are. That is, morphogenetic patterns which consist of cells over one decrease as $K$ increases. These distributions of morphogenetic patterns are due to the fact that greater values of $K$ make it more likely for GRNs to have more than two attractors so apoptosis can occur more frequently. Fig.~\ref{fig10_patterns} shows different spatial patterns of each group acquired from 20 randomly sampled simulations. The trend of the distributions in Fig.~\ref{fig9_patternDist} is visually confirmed in Fig.~\ref{fig10_patterns}. 

\begin{figure*}[htbp]
\subcapcentertrue
\subfigure[Average clustering coefficient.]{
\includegraphics[width=.3\textwidth,height=4cm]{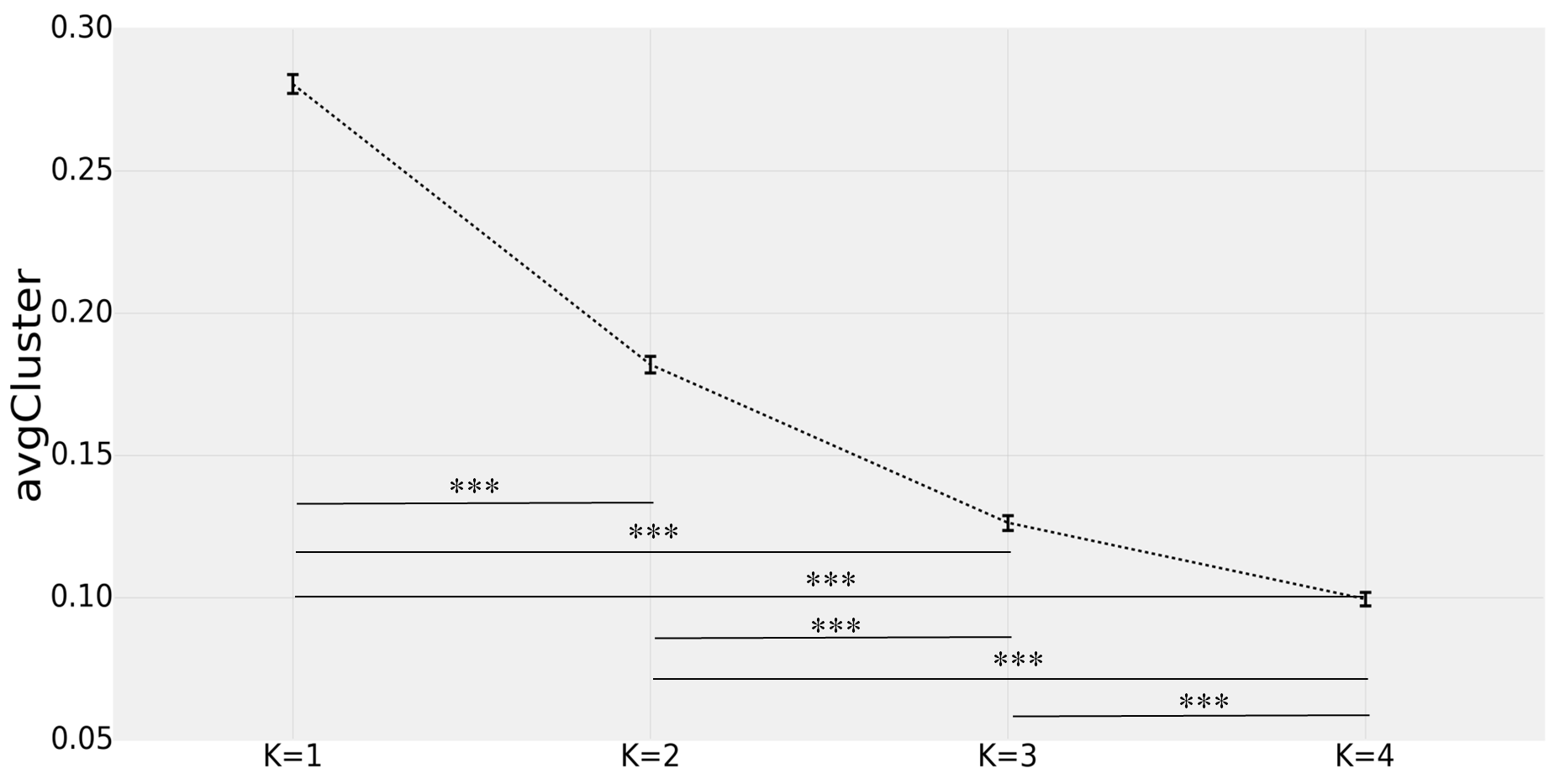}
}
\subfigure[Homogeneity of sizes of connected components.]{
\includegraphics[width=.3\textwidth,height=4cm]{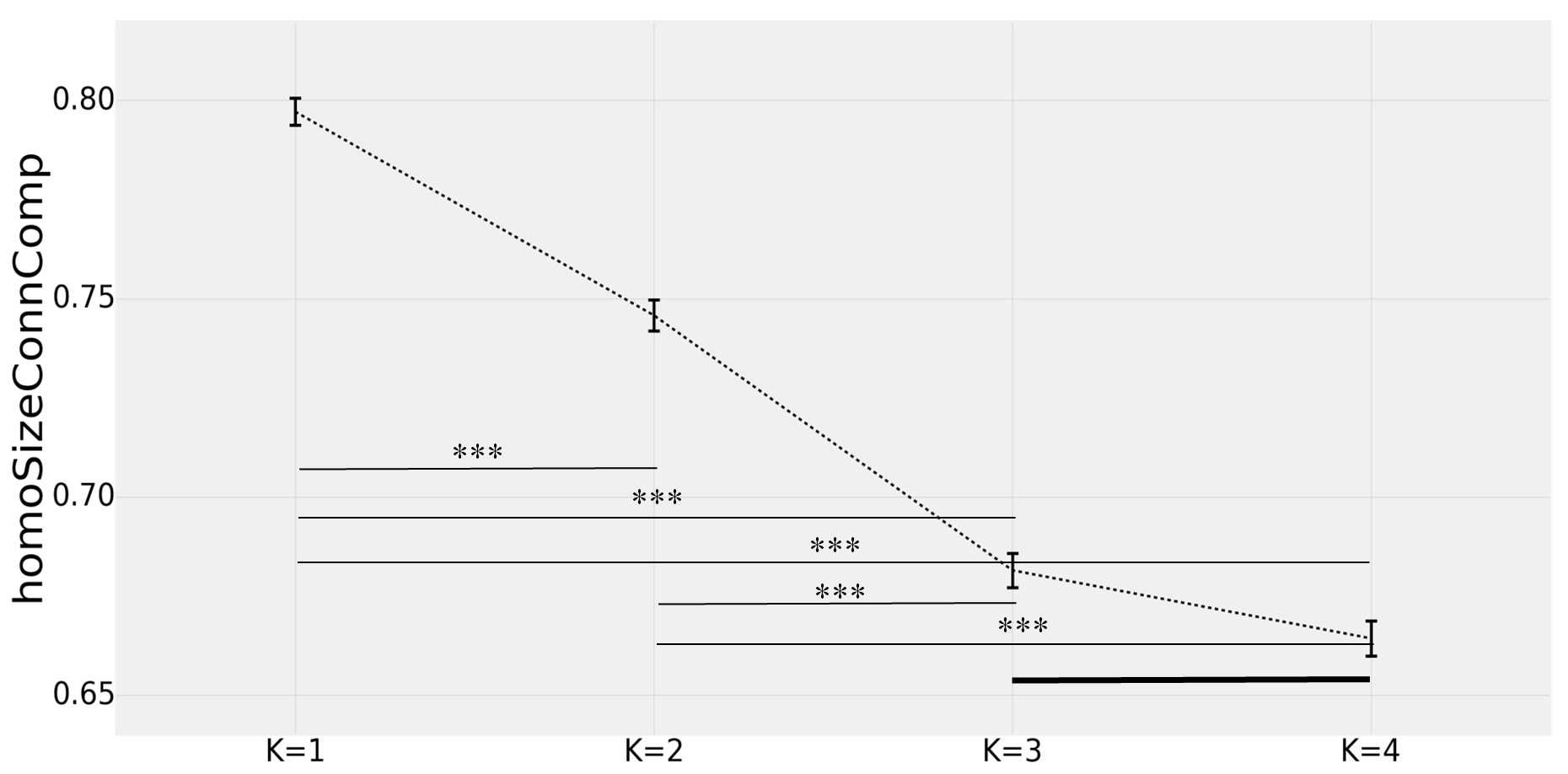}
}
\subfigure[KL divergence between pairwise distance distributions of morphologies.]{
\includegraphics[width=.3\textwidth,height=4cm]{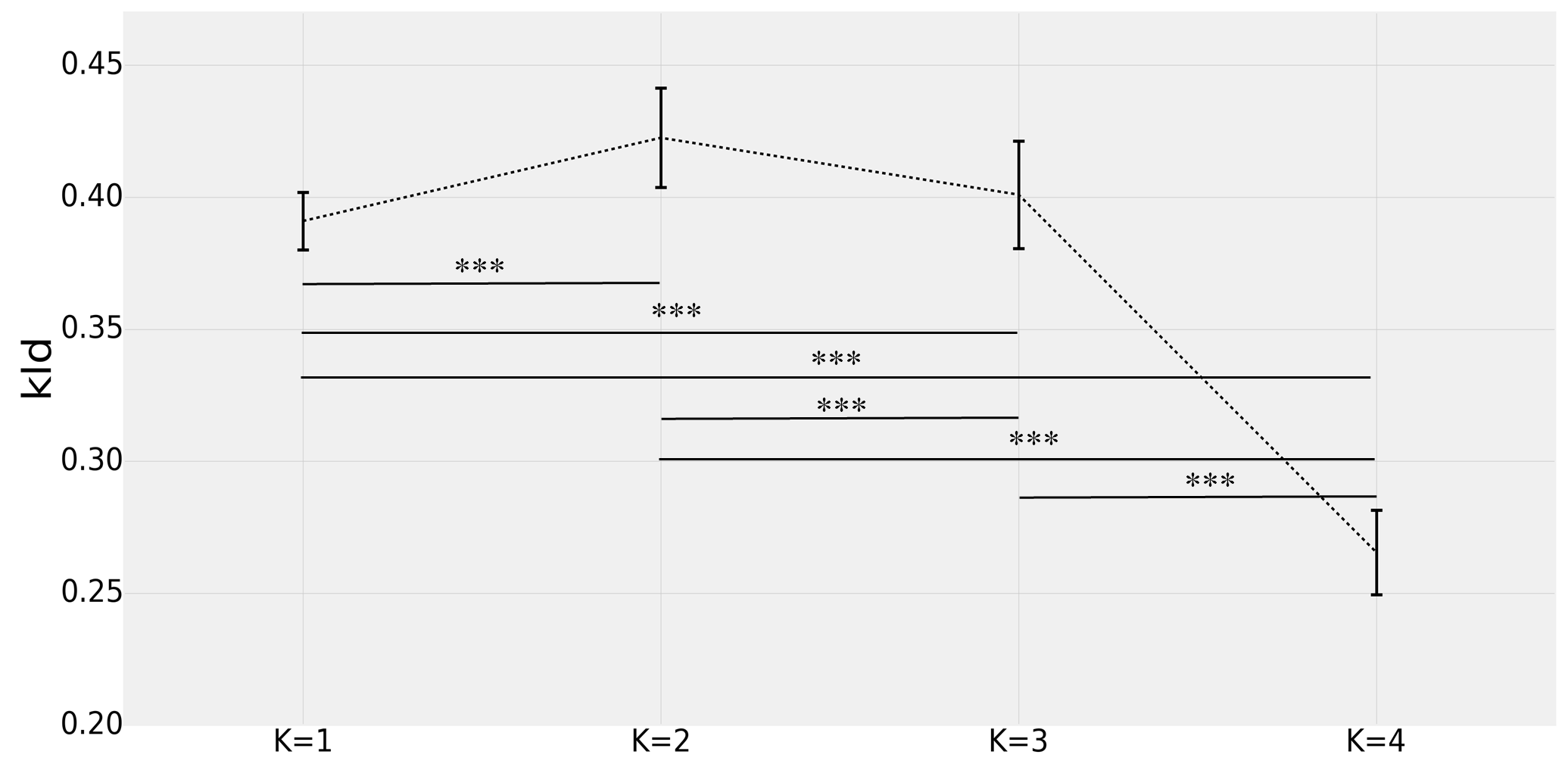}
}

\subcapcentertrue
\subfigure[Link density.]{
\includegraphics[width=.3\textwidth,height=4cm]{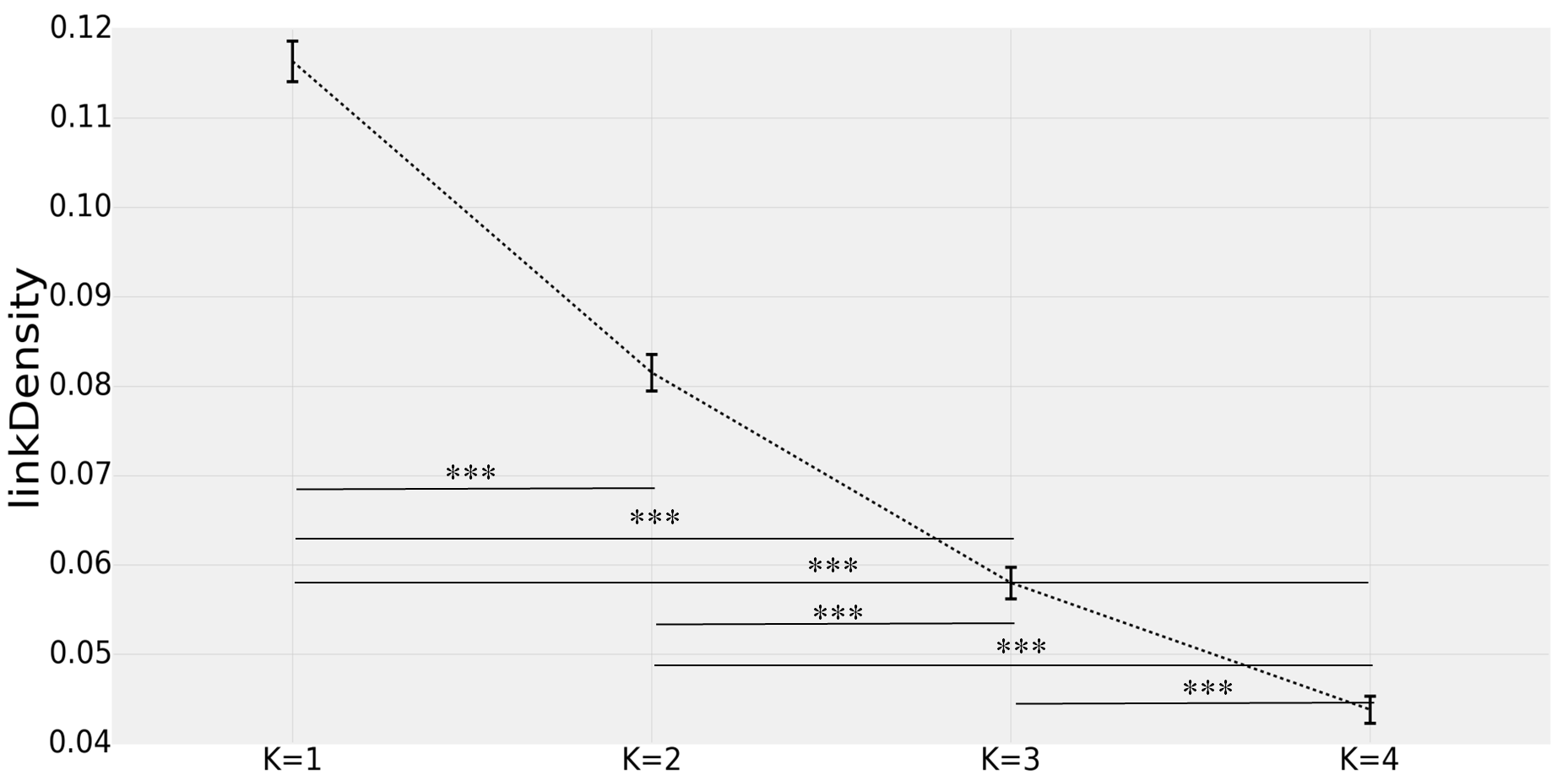}
}
\subfigure[Average distance of cells from center of mass.]{
\includegraphics[width=.3\textwidth,height=4cm]{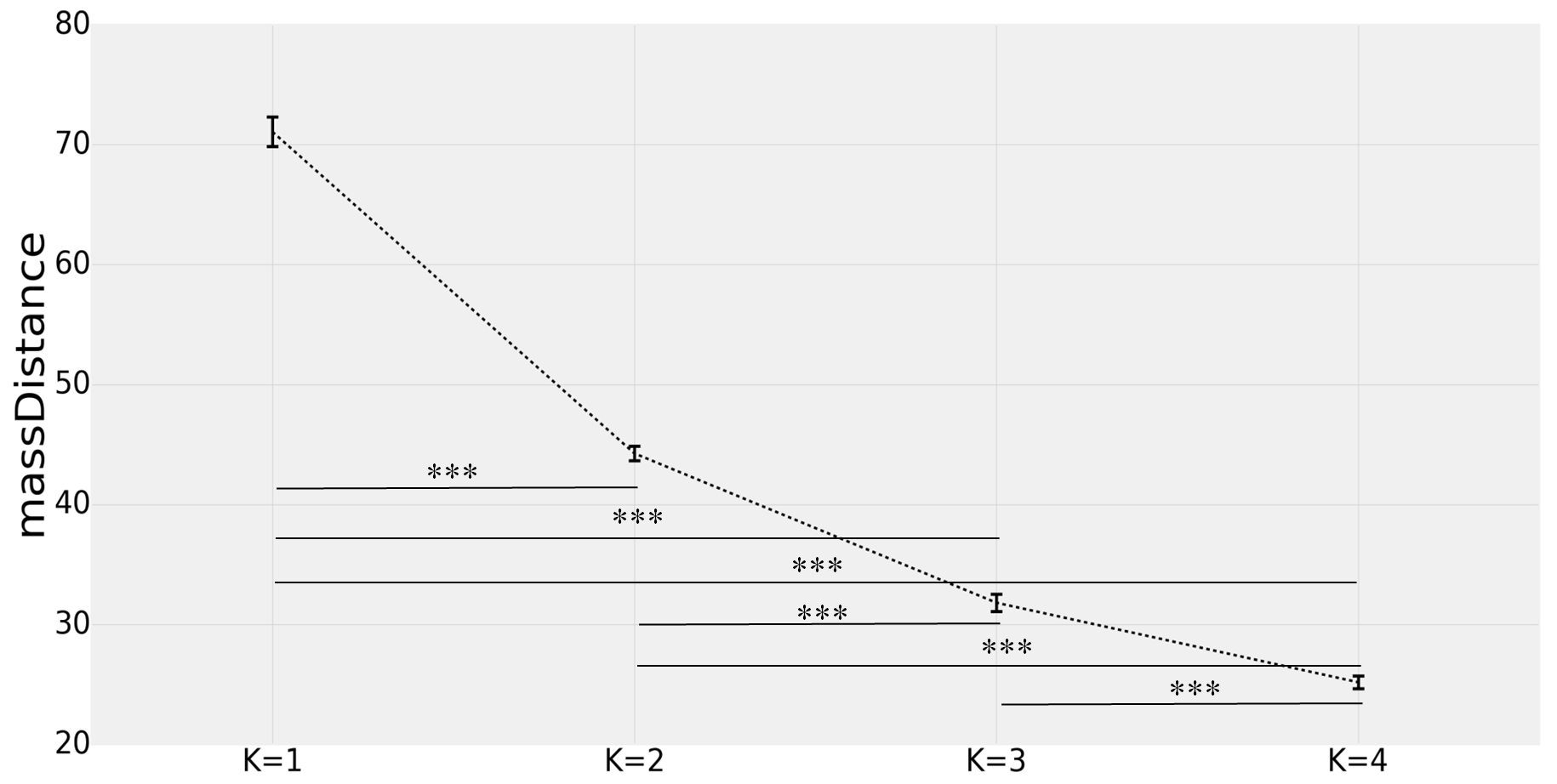}
}
\subfigure[Average size of connected components.]{
\includegraphics[width=.3\textwidth,height=4cm]{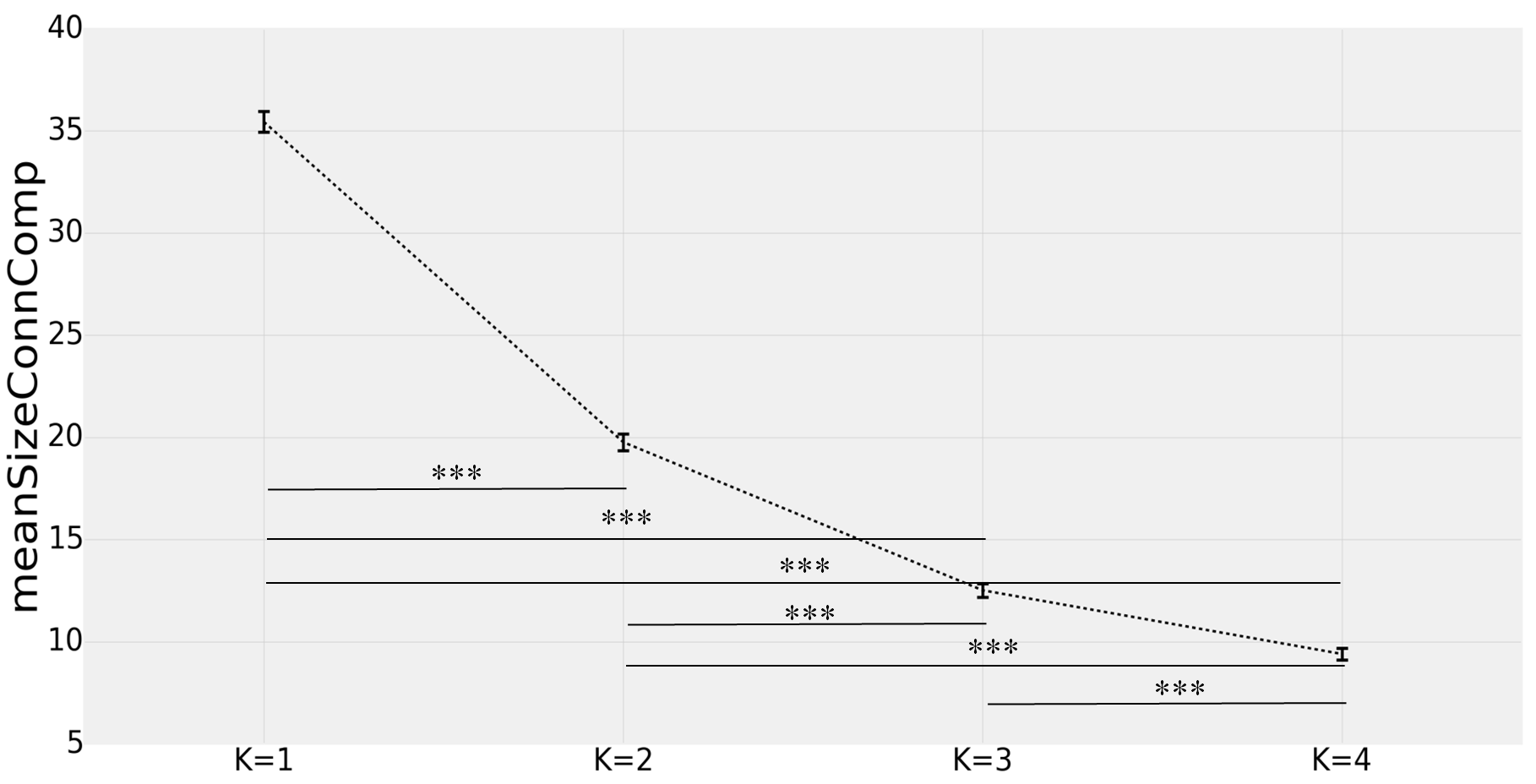}
}

\subcapcentertrue
\subfigure[Average size of connected components smaller than the largest one.]{
\includegraphics[width=.3\textwidth,height=4cm]{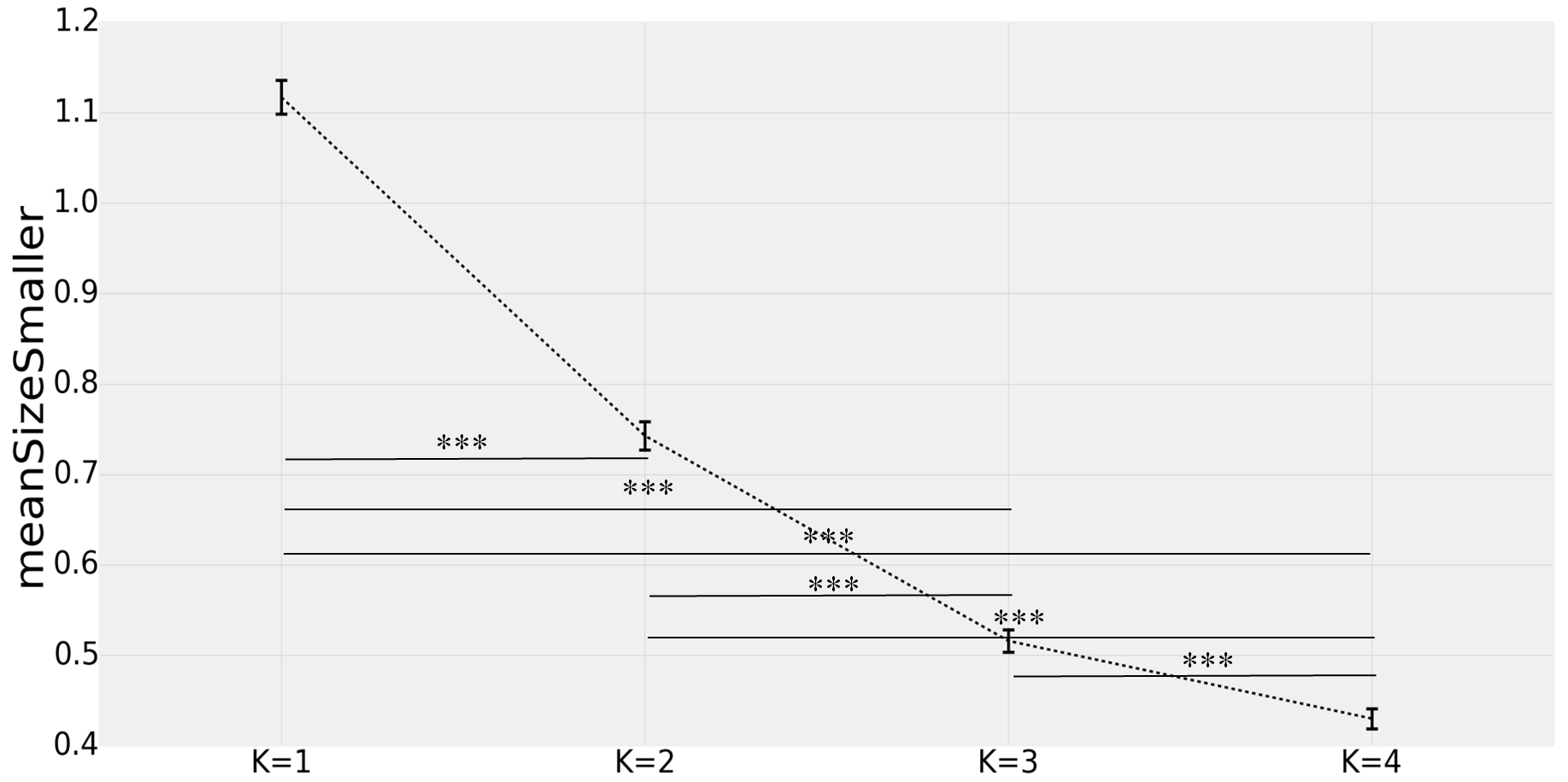}
}
\subfigure[Number of connected components.]{
\includegraphics[width=.3\textwidth,height=4cm]{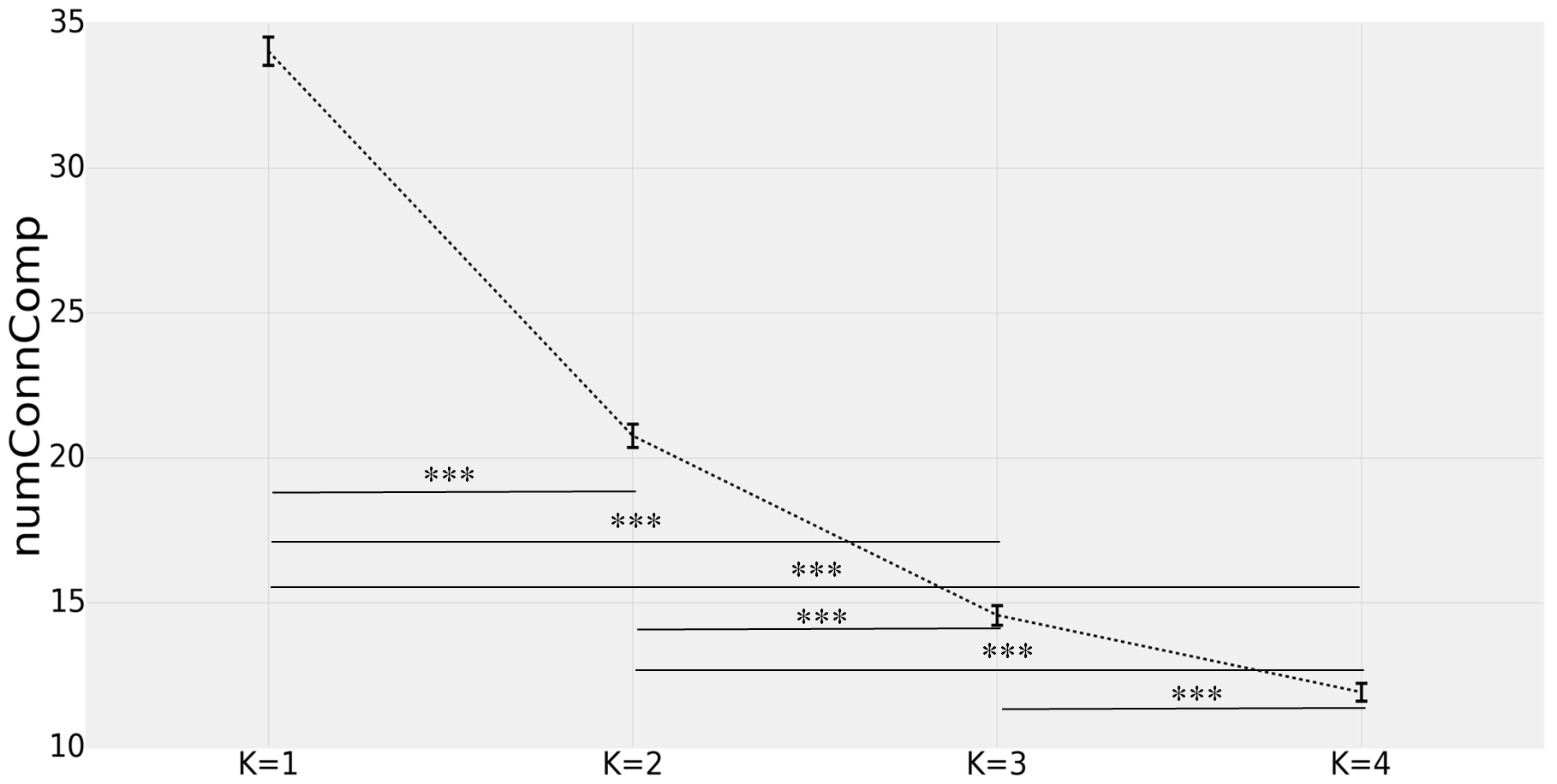}
}
\subfigure[Number of cells.]{
\includegraphics[width=.3\textwidth,height=4cm]{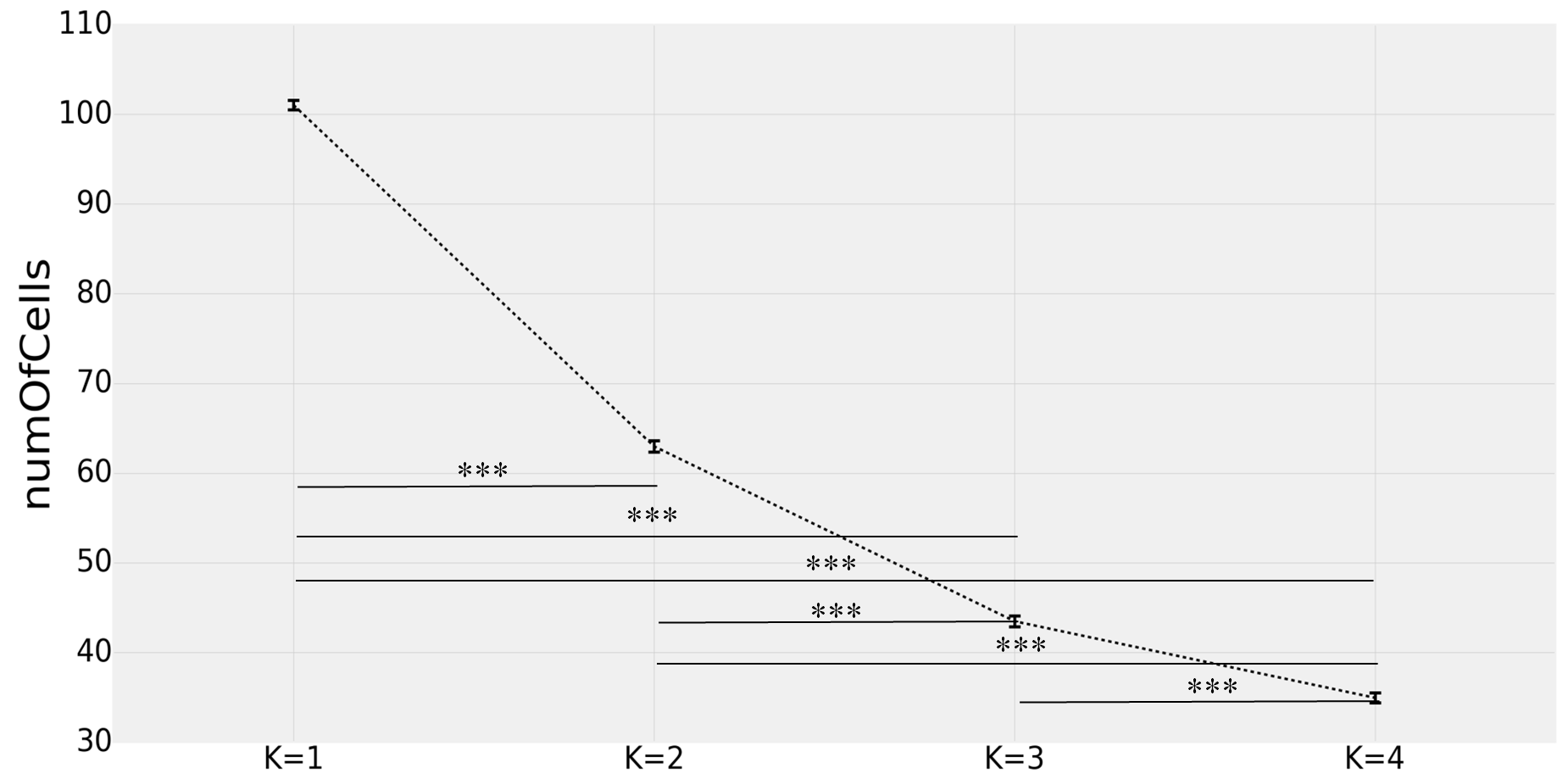}
}

\subcapcentertrue
\subfigure[Average pairwise distance.]{
\includegraphics[width=.3\textwidth,height=4cm]{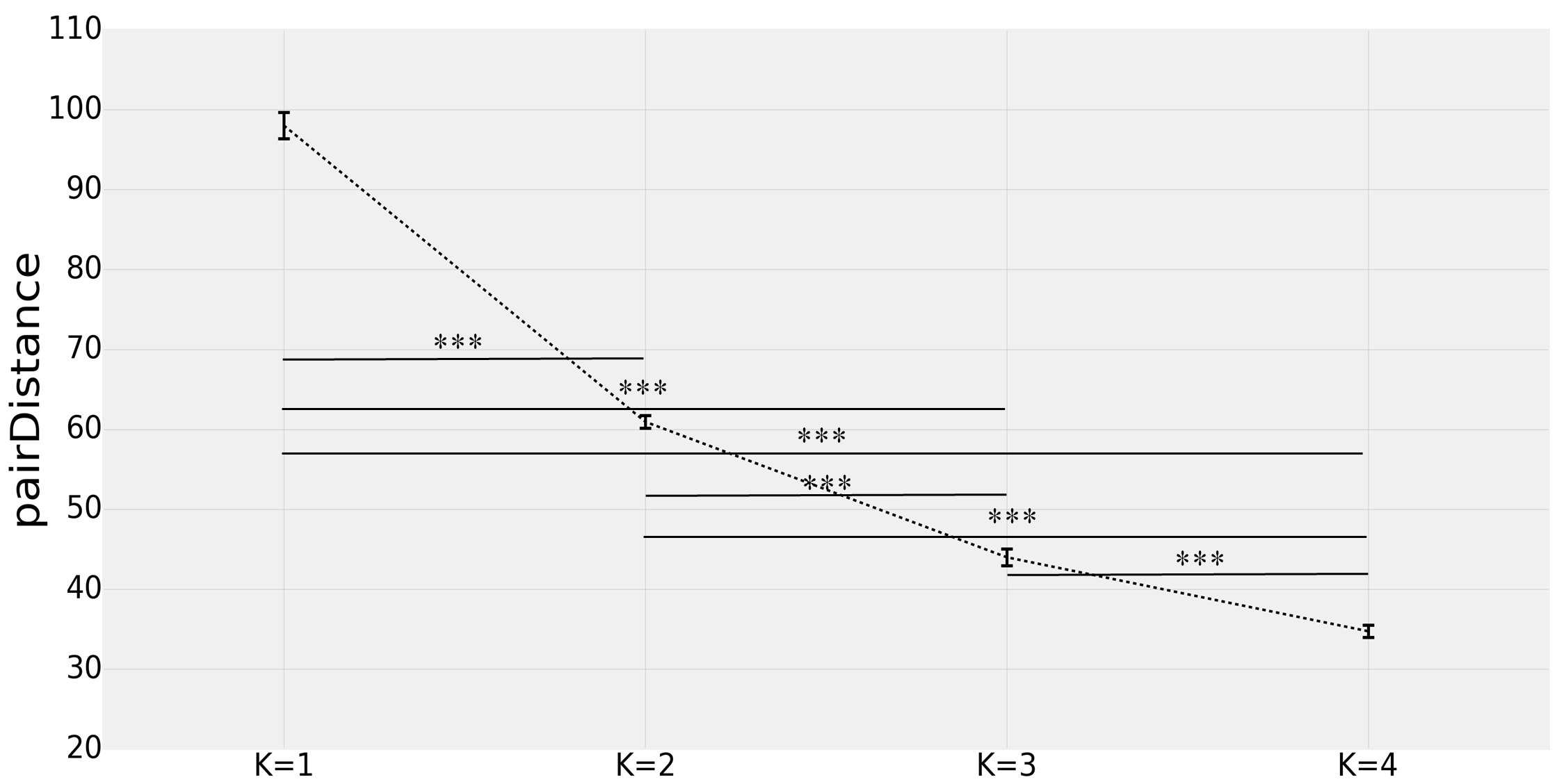}
}
\subfigure[Size of the largest connected component.]{
\includegraphics[width=.3\textwidth,height=4cm]{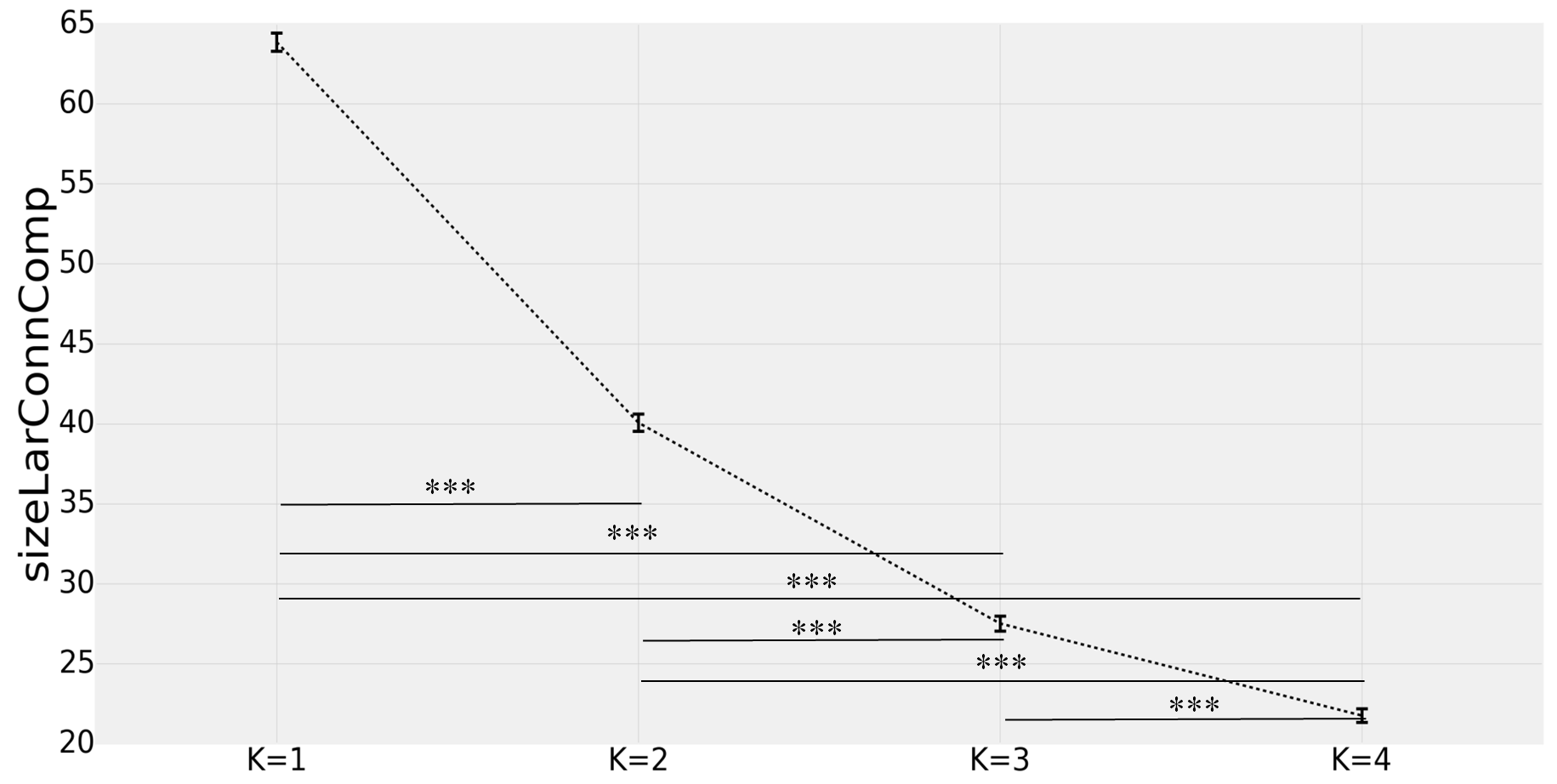}
}
\subfigure[Mutual information between different cell fates.]{
\includegraphics[width=.3\textwidth,height=4cm]{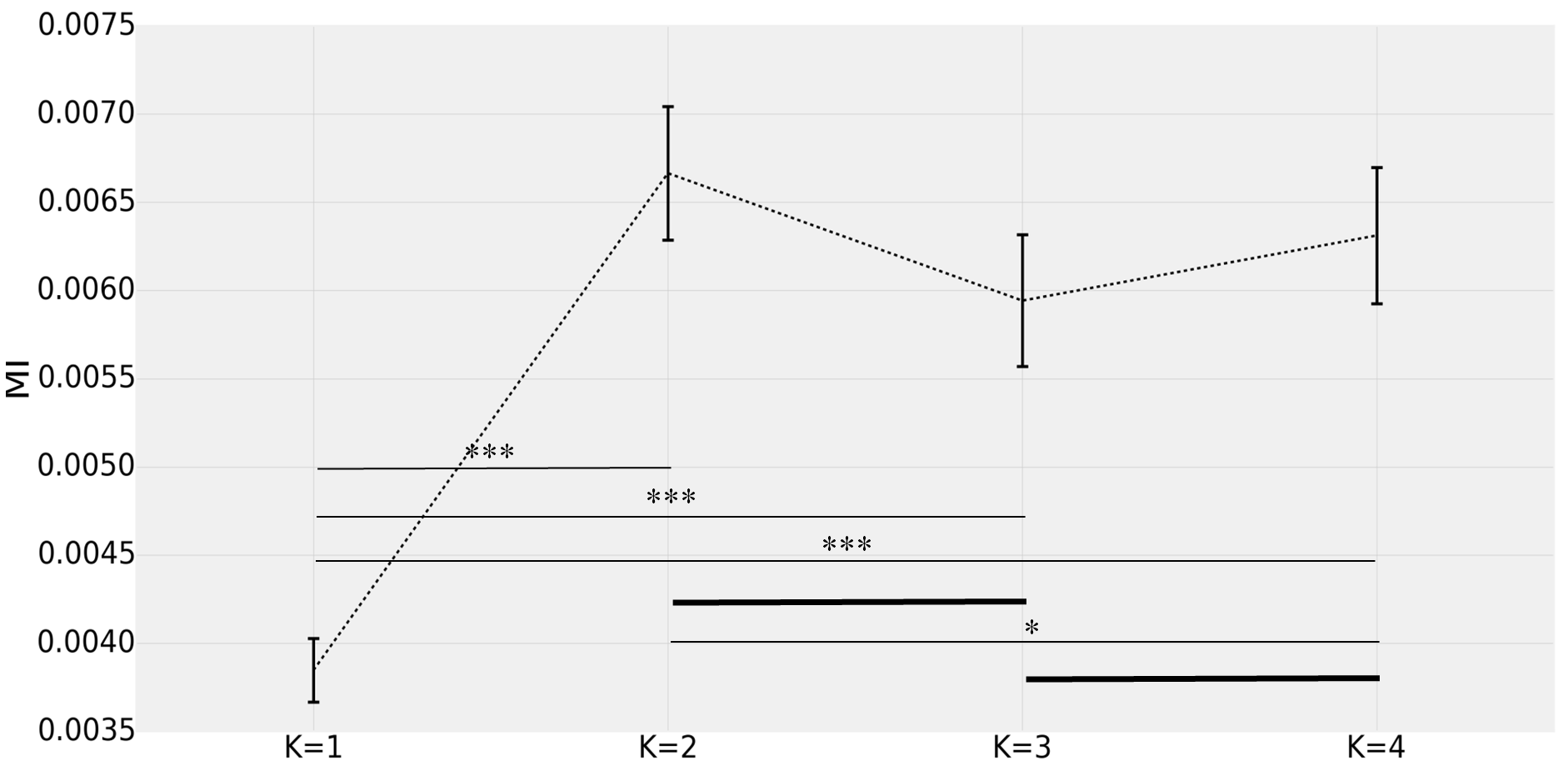}
}
\caption{Comparison of means between groups ($K = 1, 2, 3, 4$) for 12 morphological measures (Kruskal-Wallis test: $p < 2.2e-16$, Nemenyi test (post-hoc): ` ': $p <1.0$, `.': $p < 0.1$, `*': $p < 0.05$, `**': $p < 0.01$, `***': $p < 0.001$). In the case that there is no difference between two groups, a bold line without an asterisk is presented in the plot. (a) Average clustering coefficient. (b) Homogeneity of sizes of connected components. (c) KL divergence between pairwise distance distributions of morphologies. (d) Link density. (e) Average distance of cells from center of mass. (f) Average size of connected components. (g) Average size of connected components smaller than the largest one. (h) Number of connected components. (i) Number of cells. (j) Average pairwise distance. (k) Size of the largest connected component. (l) Mutual information between different cell fates.}
\label{fig11_12measures}
\end{figure*}

\begin{figure*}
\centering
\includegraphics [width=0.6\textwidth]{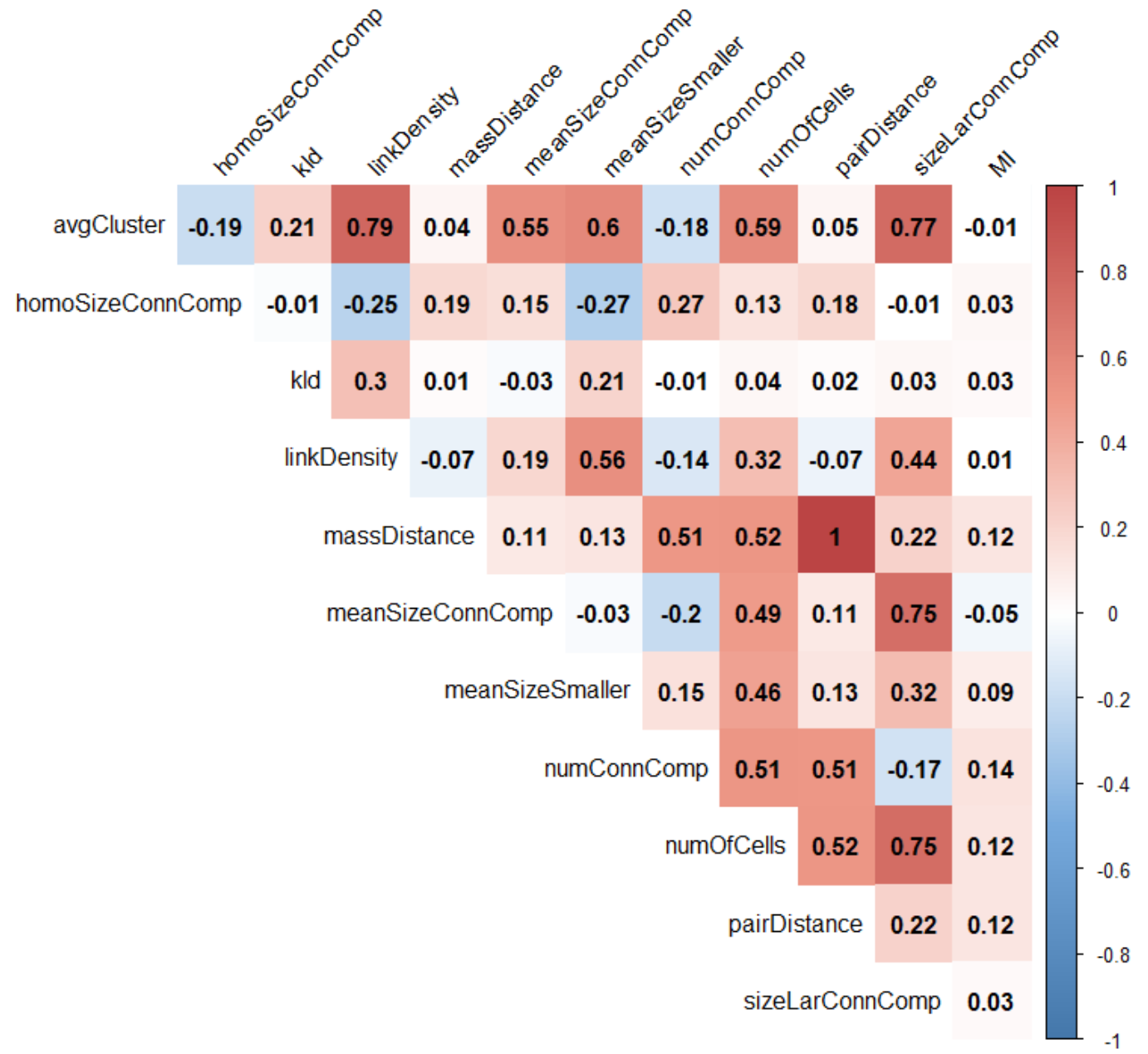}
\caption{Colored correlation matrix for 12 morphological measures.}
\label{fig12_coloredCorrMatrix} 
\end{figure*}

\begin{figure*}
\centering
\subfigure[Average basin entropy.]{
\includegraphics[width=.4\textwidth,height=5cm]{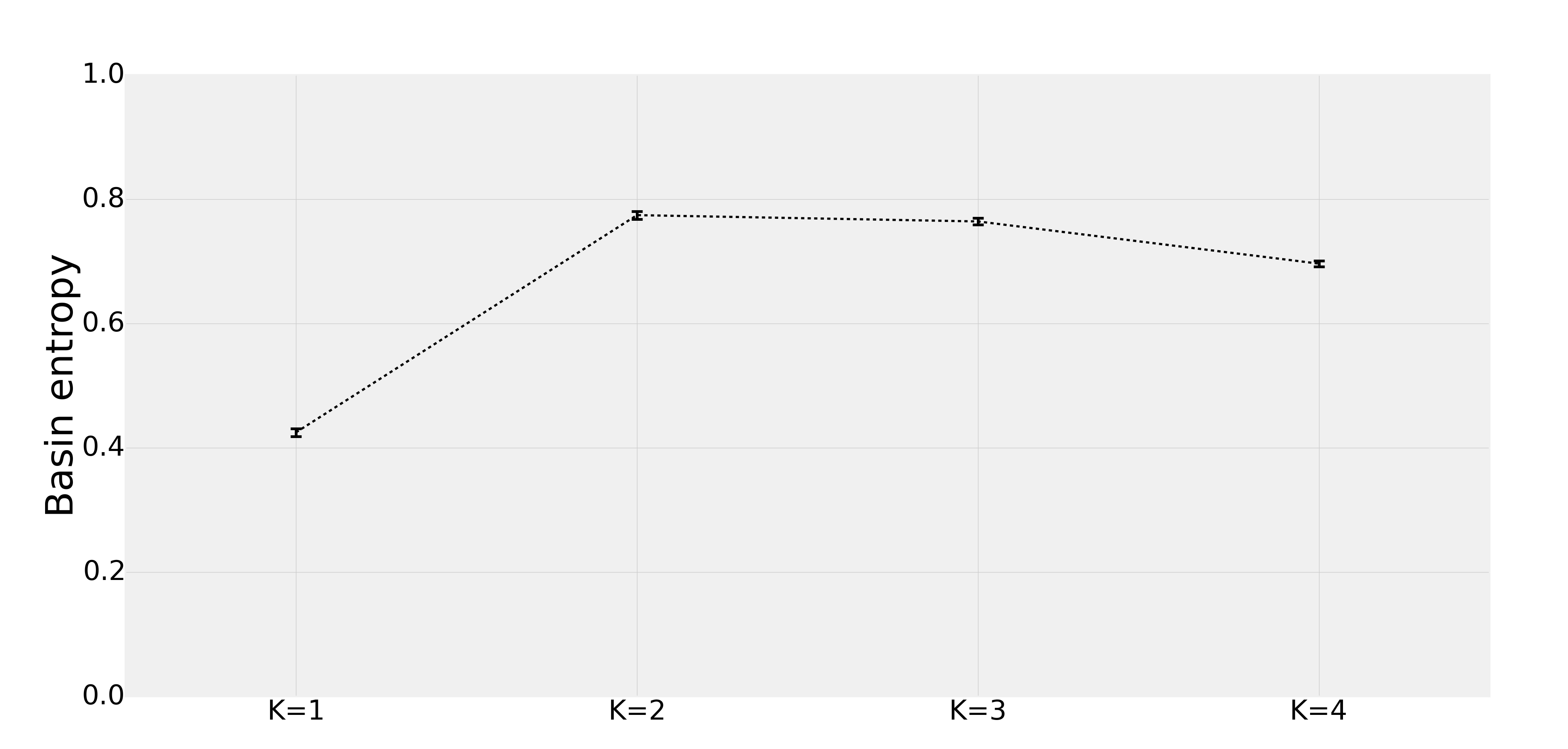}
}
\subfigure[Average state entropy of cell fates.]{
\includegraphics[width=.4\textwidth,height=5cm]{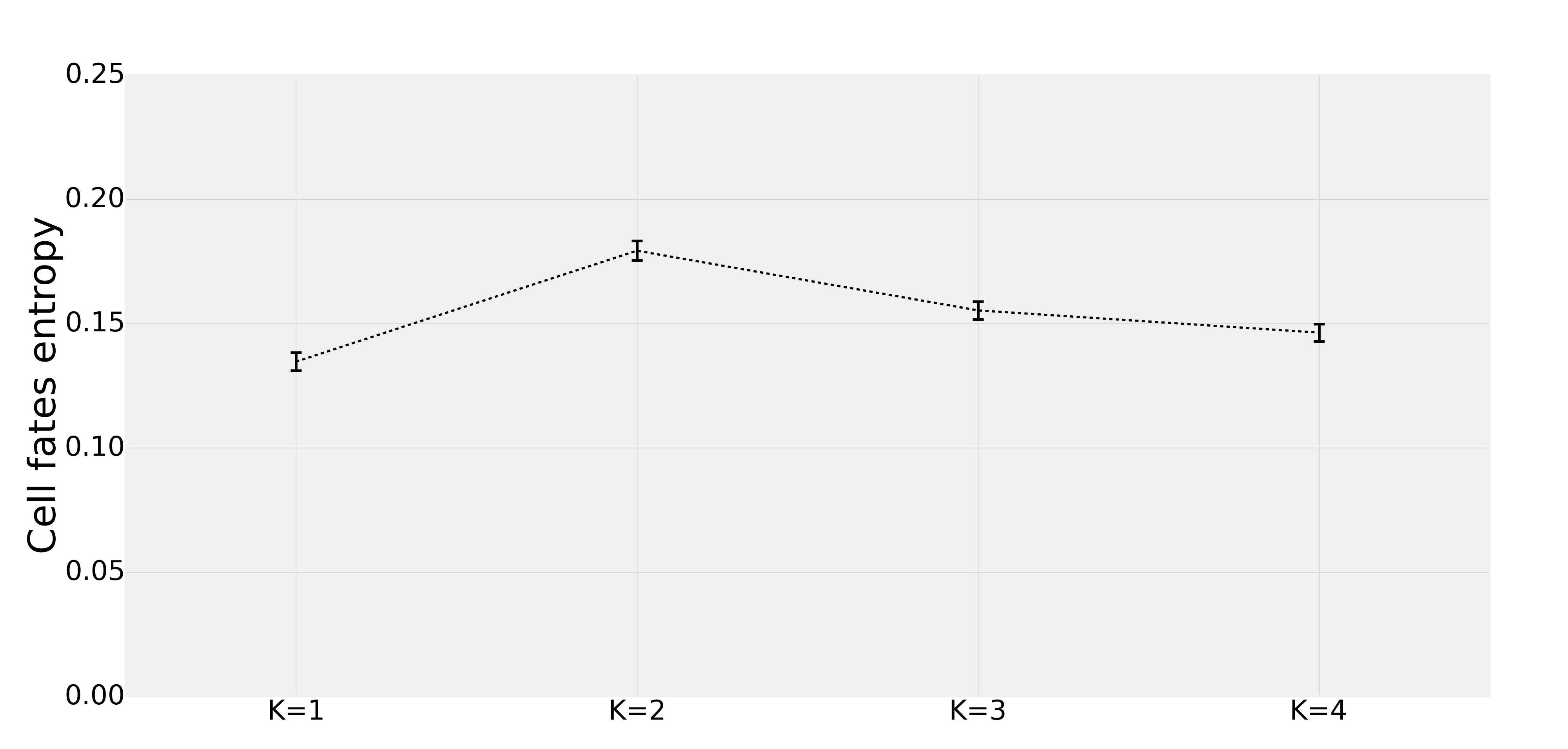}
}
\caption{Comparison of means between groups for basin and cell fates entropy. (a) Average basin entropy for $K = 1, 2, 3, 4$. (b) Average state entropy of cell fates performed in simulation at final time step for $K = 1, 2, 3, 4$.}
\label{fig13_entropies} 
\end{figure*}

Fig.~\ref{fig11_12measures} summarizes the 12 measures of spatial pattern characteristics, where Kruskal-Wallis and Nemenyi (as post-hoc analysis) tests were conducted to detect statistically significant differences among the four groups ($K = 1, 2, 3, 4$). Fig.~\ref{fig12_coloredCorrMatrix} is a correlation matrix representing correlations between the 12 measures. In the matrix, when seeing the row of \textit{numOfCells}, we found that most of the measures except for \textit{MI} and \textit{kld} were closely correlated to \textit{numOfCells}. The correlations were shown in Fig.~\ref{fig11_12measures} as well. \textit{numOfCells} decreased as $K$ increased (Fig.~\ref{fig11_12measures} (i)). This trend was also found in the measures strongly correlated with \textit{numOfCells}: \textit{avgCluster, homoSizeConnComp, linkDensity, massDistance, meanSizeConnComp, meanSizeSmaller, numConnComp, pairDistance,} and \textit{sizeLarConnComp} (Fig.~\ref{fig11_12measures} (a), (b), (d), (e), (f), (g), (h), (j), and (k)). Exceptionally, \textit{homoSizeConnComp} showed the same trend as that of \textit{numOfCells} although the correlation coefficient (i.e., 0.13) between \textit{homoSizeConnComp} and \textit{numOfCells} was small similarly to the one (i.e., 0.12) between \textit{MI} and \textit{numOfCells}. This is because the value of \textit{homoSizeConnComp} (0 $\leq$ \textit{homoSizeConnComp} $\leq$ 1) was set to 1 when there was only one connected component. In the case of \textit{single cell} in Fig.~\ref{fig9_patternDist}, the isolated single cell was considered one connected component and thus its \textit{homoSizeConnComp} was 1. These single cells resulted in the low correlation between \textit{homoSizeConnComp} and \textit{numOfCells}. Excluding \textit{single cell}, when we calculated the correlation coefficient between \textit{homoSizeConnComp} and \textit{numOfCells}, it was 0.66. We found that this strong correlation enabled \textit{homoSizeConnComp} to have the identical trend with that of \textit{numOfCells} even if the case of \textit{single cell} was included. Because the measures except for \textit{MI} and \textit{kld} are directly related to the number of components, showing the same trends as that of \textit{numOfCells}, in fact, is obvious.

Meanwhile, \textit{kld} and \textit{MI} showed different trends. The value of \textit{kld} was highest at $K=2$ (Fig.~\ref{fig11_12measures} (c)), which was counter-intuitive because the more nontrivial patterns were produced even if there were fewer patterns made of more than one cell. This result means that nontrivial morphogenetic patterns were generated most frequently when the properties of GRNs were critical. Why were more nontrivial patterns generated not at $K=1$ but at $K=2$ although $K=1$ produced more morphogenetic patterns composed of more than one cell? We can infer the reason from \textit{MI}. In Fig.~\ref{fig11_12measures} (l), the value of \textit{MI} was lowest at $K=1$ despite the greatest number of cells, which implies there existed many combinations of cells where cell states had only one cell fate, i.e., proliferation. In this case, because the same set of SMD kinetics parameters between cells ($k, l, c$ of $[$proli - proli$]$) were applied, homogeneous and circular patterns were often generated. From this, we found that cell states significantly affect the formation of nontrivial patterns.

To examine the relationship between the criticality of GRNs and the cell states of morphogenetic patterns, we measured basin entropy and cell fates entropy (Fig.~\ref{fig13_entropies}). For both of them, the average values were largest at $K=2$. It indicates that basins of attraction where cell fates were randomly assigned were most evenly distributed at $K=2$, which made the expressions of different cell fates maximally balanced. These trends of basin entropy and cell fates entropy matched nicely with the \textit{kld} measure. This implies that the maximally balanced expressions of the cell fates enabled the different parameters of SMD kinetics to be most evenly applied to cells, which finally produced nontrivial patterns most frequently at $K=2$.

\section{Conclusions}
In this study, we proposed new GRN-based morphogenetic systems using Kauffman's $NK$ RBNs and SMD kinetics to show self-organized spatial patterns during the developmental process. We simulated the model varying the properties of GRNs from ordered ($K=1$), through critical ($K=2$), to chaotic ($K=3, 4$) regimes. The simulations showed that nontrivial morphogenetic patterns were produced most frequently in morphogenetic systems with critical GRNs. Our findings indicate that the criticality of GRNs plays an important role in facilitating the formation of nontrivial morphogenetic patterns in GRN-based morphogenetic systems. 

The pattern formation in our morphogenetic systems can be interpreted as morphogenesis of multicellular organisms in the biological perspective. Not simple patterns such as one single cell or homogeneous and circular patterns but nontrivial patterns maximally emerged at $K=2$. Biologically, specific functions of cells are closely related to their complex structures \cite{lowe1960relation}. Thus, it is important to understand how the complex shapes had emerged. In our model, the facilitation of nontrivial pattern formation at $K=2$ may shed light on the morphogenesis of highly structured tissues or organs of living organisms. 

The present study has several limitations. First, the effect of criticality of GRNs in the process of cell-cell interactions has not been thoroughly explored. We measured basin and cell fates entropy to reveal the relationship between the criticality of GRNs and nontrivial pattern formation, which does not fully account for the trend of \textit{kld} ($K=2 > K=3 > K=1 > K=4$). This is because we only focused on morphogenetic patterns acquired at the final time step of the simulation. Therefore, for further study, we will look into the spatial and temporal distribution of cells during the developmental process, tracking the whole process that cell fates are determined by the interactions of neighboring cells from the initial time step to the final time step.

Second, an evolutionary process of GRNs is not included. We simulated our model without considering the change of GRNs in an evolutionary sense. Because GRNs can be changed by mutations caused by internal or external factors, we plan to introduce perturbations such as adding, deleting, or switching links to GRNs, and investigate if the role of criticality of GRNs can be maintained. Furthermore, using complexity measures, we will compare the complexity of the morphogenetic patterns per $K$ to find which regime of GRNs is evolutionarily optimized.

Finally, our model remains highly artificial and is limited in offering biologically realistic predictions. We used artificial RBNs as GRNs of our model and SMD kinetics for cellular movements, which were not constructed faithfully to real biology. Thus, by using empirically obtained biological Boolean networks \cite{li2004yeast, chaves2006methods, alvarez2007gene} and the mechanisms of morphogenetic cell movements, we will suggest more  biologically improved model and explore its potential applications.

\section*{Acknowledgment}
This material is based upon work supported by the US National Science Foundation under Grant No. 1319152.

\bibliographystyle{IEEEtran}
\bibliography{kim-sayama}

\end{document}